\documentclass[a4paper,cite,11pt,psnfss]{article}

\usepackage[T1]{fontenc}
\usepackage[latin1]{inputenc}
\usepackage{comment}
\usepackage{ulem}
\usepackage{adjustbox}
\usepackage{appendix}
\usepackage{framed,epsfig,pgfplots,amsmath}
\usepackage{tikz}
\usepackage{tikz-feynhand}
\usetikzlibrary{snakes,arrows,shapes,positioning,automata,backgrounds,calc,er,patterns}
\usepackage[usenames,dvipsnames,tree]{pstricks}

\usepackage{graphicx,color,pst-plot}

\usepackage{latexsym,amsmath,amsfonts,amssymb,amstext,mathrsfs,upgreek}

\allowdisplaybreaks[1]

\begin{document}

\begin{titlepage}
\begin{flushright}
\end{flushright}
\begin{flushright}
\end{flushright}

\vfill

\begin{center}

{\Large\bf The Double Box and Hexagon Conformal Feynman Integrals}

\vfill

{\bf B. Ananthanarayan$^a$, Sumit Banik$^a$, Samuel Friot$^{b,c,}$\footnote{Correspondence: samuel.friot@universite-paris-saclay.fr} and Shayan Ghosh$^d$}\\[1cm]
{$^a$ Centre for High Energy Physics, Indian Institute of Science, \\
Bangalore-560012, Karnataka, India}\\[0.5cm]
{$^b$ Universit\'e Paris-Saclay, CNRS/IN2P3, IJCLab, 91405 Orsay, France } \\[0.5cm]
{$^c$ Univ Lyon, Universit\'e Claude Bernard Lyon 1, CNRS/IN2P3, \\
 IP2I Lyon, UMR 5822, F-69622, Villeurbanne, France}\\[0.5cm]
{$^d$ Helmholtz-Institut f\"ur Strahlen- und Kernphysik \& Bethe Center for Theoretical Physics, Universit\"at Bonn, D-53115 Bonn, Germany} \\
\end{center}
\vfill

\begin{abstract}
The off-shell massless six-point double box and hexagon conformal Feynman integrals with generic propagator powers are expressed in terms of linear combinations of multiple hypergeometric series of the generalized Horn type. These results are derived from 9-fold Mellin-Barnes representations obtained from their dual conformal Feynman parameter representations.
The individual terms in the presented expressions satisfy the differential equation that relates the double box in $D$ dimensions to the hexagon in $D+2$ dimensions.

\end{abstract}

\end{titlepage}

\section{Introduction}
It was noted in a recent work \cite{Loebbert:2019vcj} that there is a close connection between the Mellin-Barnes (MB) computational technique and the constraints coming from a recently discovered Yangian symmetry of conformal Feynman integrals \cite{Chicherin:2017cns,Chicherin:2017frs}. The authors of \cite{Loebbert:2019vcj} argue that the MB method can be used to generate Yangian invariants as sums of residues. However they also point out, for the important cases of the analytically unknown off-shell massless six point $D$-dimensional double box integral and related hexagon, that there are complications, both in the MB and Yangian approaches, in identifying the latter integrals as particular combinations of these invariants. Thus, the resolution of these remains a challenging and unsolved problem.

To make a breakthrough in this context, we have extended the computational technique of \cite{TZ,Friot:2011ic} in order to deal with $N$-fold MB integrals.
 This allows us to solve the problem above from the MB side, by extracting series representations of both the double box and hexagon Feynman integrals with generic propagator powers from their $9$-fold MB representations in a systematic way. The results are given in terms of linear combinations of multiple hypergeometric series of the generalized Horn type. We emphasize that the derivation of these linear combinations does not require convergence considerations. Regarding the large number of possible series to build the linear combinations, 2530 in the hexagon case and 4834 for the double box,
this amounts to a distinct advantage compared to the corresponding situation in the Yangian bootstrap approach of \cite{Loebbert:2019vcj} where the convergence properties of the involved series seem to be a necessary external input.

For each of the 9-fold MB integrals, one can obtain hundreds of different linear combinations  which are analytic continuations of one another converging in different regions of the 9-dimensional space of the cross ratios that enter the expressions. Obviously the full set of these linear combinations cannot be presented here. We therefore focus on one example of such expressions for both the double box and hexagon, which is explicitly given in a supplemental material to this letter. The individual terms in the presented expressions satisfy the differential equation that relates the double box in $D$ dimensions to the hexagon in $D+2$ dimensions. In addition to solving the thus far unsolved problem of evaluating these complicated Feynman integrals, it is likely that such an approach may yield insights for the solution of the corresponding Yangian constraints.

\section{Mellin-Barnes representations of the double box and hexagon}
 
We follow in our computations the notational conventions of \cite{Loebbert:2019vcj} where the six point off-shell massless double box with generic propagator powers (see Fig. 1 left) is written, in dual momentum space, as 
\begin{equation}
I_{3,3}=\int\frac{\text{d}^Dx_0\text{d}^Dx_{0'}}{x_{10}^{2a}x_{20}^{2b}x_{30}^{2c}x_{00'}^{2\ell}x_{40'}^{2d}x_{50'}^{2e}x_{60'}^{2f}}=V_{3,3}\phi_{3,3}
\end{equation}
with $\phi_{3,3}$ a conformally invariant function of nine cross ratios ($u_i$, $i=1,...,9$) and $V_{3,3}$ is the prefactor
\begin{equation}
V_{3,3}=x_{13}^{2\ell-D}x_{14}^{D-2\ell}x_{15}^{-2d-2e}x_{16}^{2d+2e-2a}x_{26}^{-2b}x_{36}^{D-2c-2\ell}x_{46}^{2\ell-2d-D}x_{56}^{2d}
\end{equation}
The dual conformal Feynman parameter representation of the double box is given in \cite{Loebbert:2019vcj}, where
\begin{equation}
\phi_{3,3}(u_1,...,u_9,D)=Q_{3,3}\int_0^\infty \text{d}\beta_2\text{d}\beta_3\text{d}\beta_4\text{d}\beta_5\frac{\beta_2^{b-1}\beta_3^{c-1}\beta_4^{d-1}\beta_5^{e-1}}{X_2^{D/2-\ell}Y^{D/2-f}Z_4^f}
\label{FPDB}
\end{equation}
with
\begin{equation}
Q_{3,3}=\frac{\pi^{D}\Gamma(D/2-\ell)\Gamma(D/2-f)}{\Gamma(a)\Gamma(b)\Gamma(c)\Gamma(l)\Gamma(d)\Gamma(e)}
\end{equation}
\begin{equation}
X_2=\beta_2u_6u_9+\beta_3u_9+\beta_2\beta_3u_1u_2u_3u_9
\end{equation}
\begin{equation}
Y=u_8X_2+\beta_4u_9+\beta_2\beta_4u_2u_3u_9+\beta_3\beta_4u_2u_3u_4u_5u_9+\beta_5+\beta_2\beta_5u_3+\beta_3\beta_5u_3u_5+\beta_4\beta_5u_3u_5u_7
\end{equation}
and
\begin{equation}
Z_4=1+\beta_2+\beta_3+\beta_4+\beta_5
\end{equation}

In the expressions above, $x^\mu_{ij}=x^\mu_i-x^\mu_j$ and the momenta are related to their duals by $p_j^\mu=x_j^\mu-x_{j+1}^\mu$.
Note that the conformal constraints are $a+b+c+\ell=D$ and $d+e+f+\ell=D$.

\begin{figure}
\begin{center}
\includegraphics[scale=0.03]{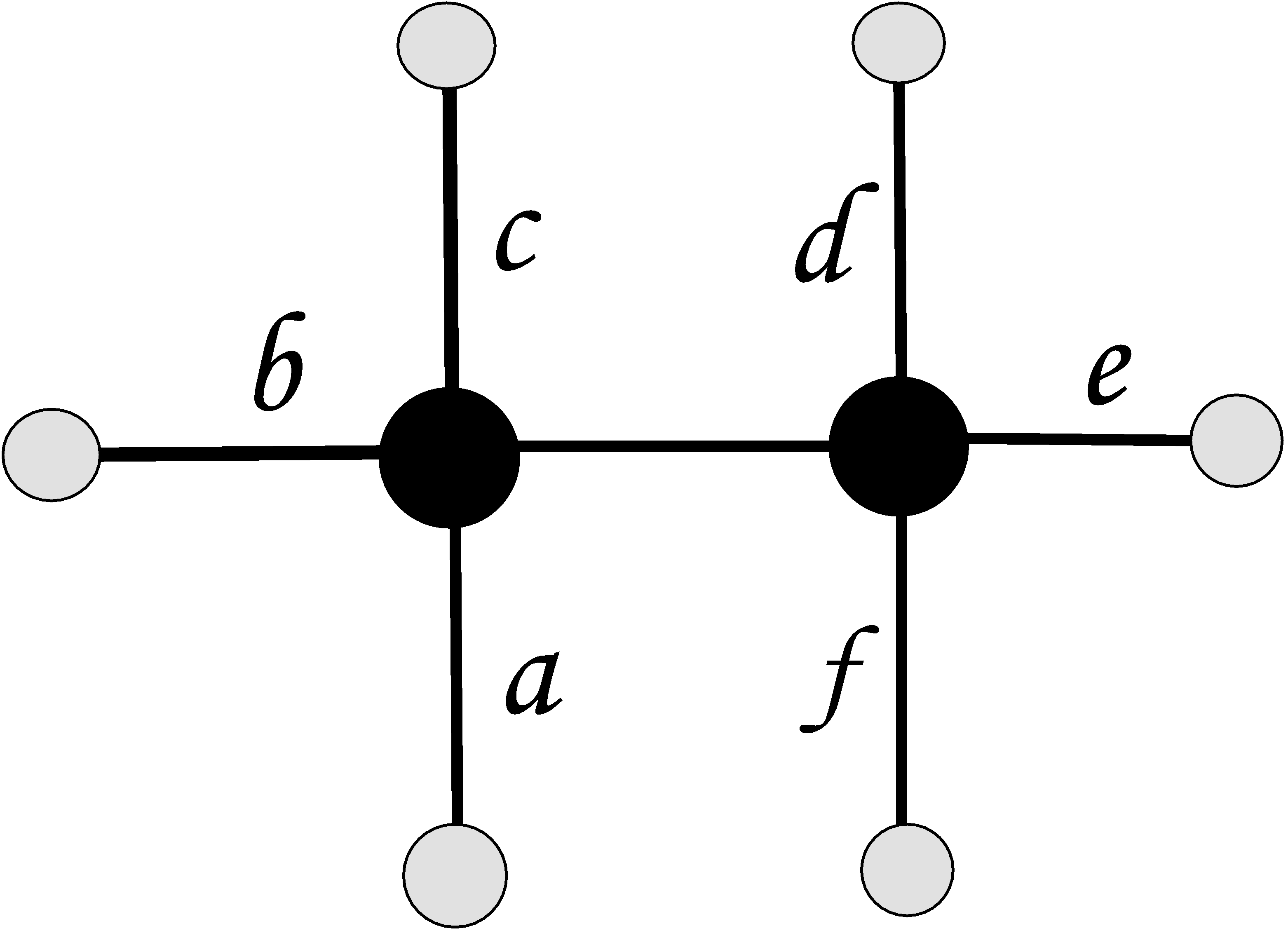}\hspace{2cm}
\includegraphics[scale=0.03]{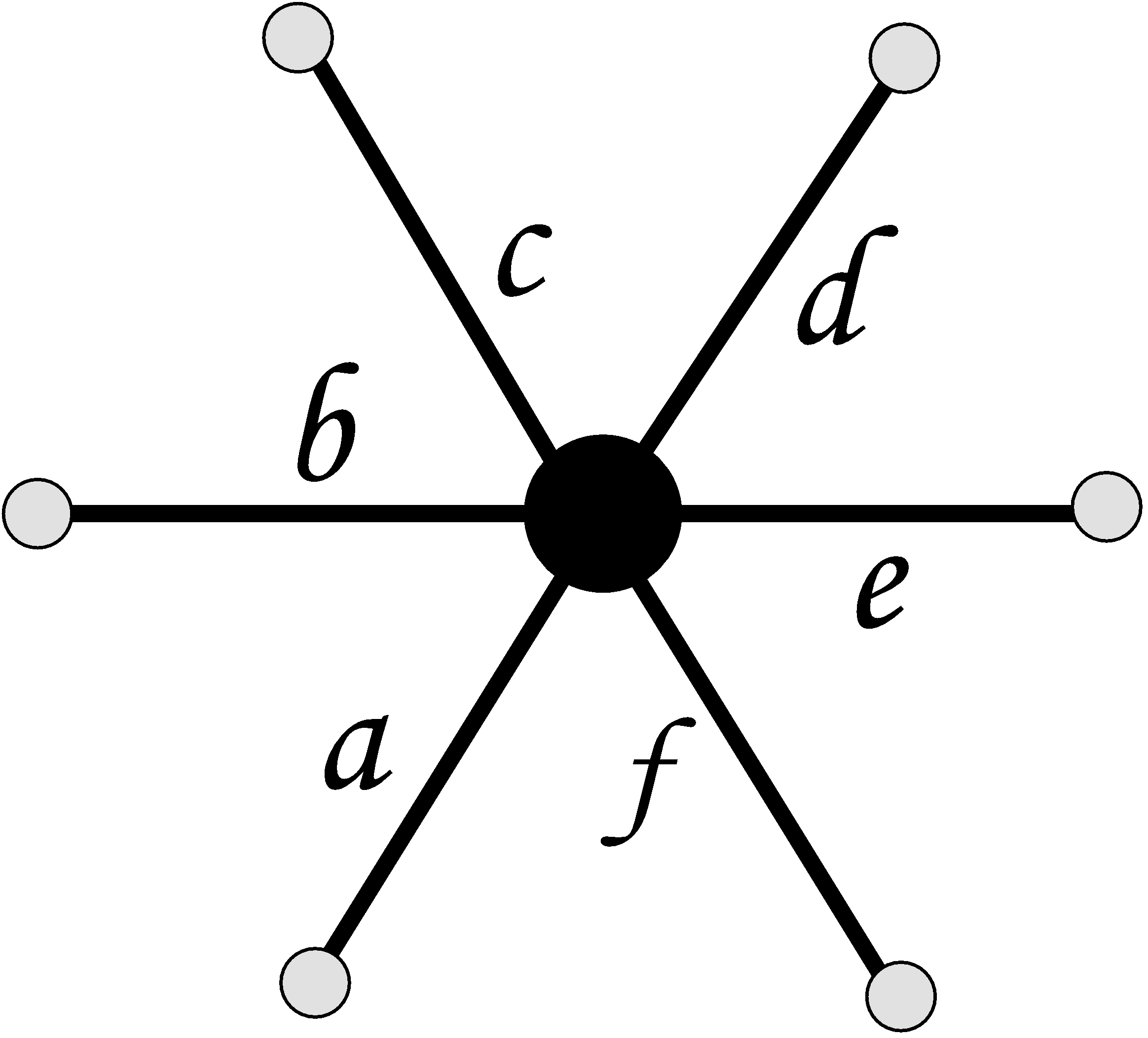}
\end{center}
\caption{The double box and hexagon integrals.}
\end{figure}
  \bigskip

A similar analysis yields the hexagon integral (see Fig. 1 right) as
\begin{equation}
I_{6}=\int\frac{\text{d}^Dx_0}{x_{10}^{2a}x_{20}^{2b}x_{30}^{2c}x_{40}^{2d}x_{50}^{2e}x_{60}^{2f}}=V_6\phi_6
\end{equation}
where \cite{Loebbert:2019vcj}
\begin{equation}
\phi_6(u_1,...,u_9,D)=Q_6\int_0^\infty \text{d}\beta_2\text{d}\beta_3\text{d}\beta_4\text{d}\beta_5\frac{\beta_2^{b-1}\beta_3^{c-1}\beta_4^{d-1}\beta_5^{e-1}}{Y^{D/2-f}Z_4^f}
\label{FPhex}
\end{equation}
with
\begin{equation}
Q_6=\frac{\pi^{D/2}\Gamma(D/2-f)}{\Gamma(a)\Gamma(b)\Gamma(c)\Gamma(d)\Gamma(e)}
\end{equation}
and $V_6=x_{15}^{2f-D}x_{16}^{D-2a-2f}x_{26}^{-2b}x_{36}^{-2c}x_{46}^{-2d}x_{56}^{D-2e-2f}$. 

In this case the conformal constraint is $a+b+c+d+e+f=D$.

\bigskip

It is straightforward to derive the MB representations  of the double box and hexagon from the Feynman parameterizations given in Eqs. (\ref{FPDB}) and (\ref{FPhex}) which can be written as
\begin{align}
\phi_{3,3}&=\frac{Q_{3,3}\ u_8^{D/2-\ell}}{\Gamma(D/2-f)\Gamma(f)}\frac{1}{(2i \pi)^9}\int_{-i\infty}^{+i\infty}\text{d}z_1...\int_{-i\infty}^{+i\infty}\text{d}z_9\Pi_{i=1}^9\left(w_i^{z_i}\Gamma(-z_i)\right)\Gamma(D-f-\ell+z_1+...+z_9)\nonumber\\
&\times\Gamma(b+z_1+z_5+z_8+z_9)\Gamma(c+z_2+z_6+z_7+z_8)
\Gamma(d+z_3+...+z_6)\nonumber\\
&\times\Gamma(e+f+\ell-D-z_4-...-z_9)\Gamma(-b-c-d-e+D-\ell-z_1-z_2-z_3-z_5-z_6-z_8)\nonumber\\
&\times\frac{\Gamma(\ell-D/2-z_7-z_8-z_9)}{\Gamma(-z_7-z_8-z_9)}
\label{DB}
\end{align}
and
\begin{align}
\phi_6&=\frac{Q_6}{\Gamma(D/2-f)\Gamma(f)}\frac{1}{(2i \pi)^9}\int_{-i\infty}^{+i\infty}\text{d}z_1...\int_{-i\infty}^{+i\infty}\text{d}z_9\Pi_{i=1}^9\left(w_i^{z_i}\Gamma(-z_i)\right)\Gamma(D/2-f+z_1+...+z_9)\nonumber\\
&\times\Gamma(b+z_1+z_5+z_8+z_9)\Gamma(c+z_2+z_6+z_7+z_8)
\Gamma(d+z_3+...+z_6)\nonumber\\
&\times\Gamma(e+f-D/2-z_4-...-z_9)\Gamma(-b-c-d-e+D/2-z_1-z_2-z_3-z_5-z_6-z_8)
\label{hex}
\end{align}
where the $w_i$ are combinations of the $u_i$ (see Eq. (A20) in \cite{Loebbert:2019vcj}).

The two integrals in Eqs.(\ref{DB}) and (\ref{hex}) belong to a class of MB integrals for which several series representations of the latter coexist, converging in various regions of the cross ratios' 9-dimensional space. 

As a final remark, one should note that the conformally invariant functions of the double box in $D$ dimensions and hexagon in $D+2$ dimensions are related by the differential equation
\begin{equation}
\partial_{u_8}\phi_{3,3}(u_1,...,u_9,D)=-\frac{\pi^{D/2-1}}{\Gamma(l)}\phi_6(u_1,...,u_9,D+2)
\label{diffeq}
\end{equation}
when $D/2-l=1$,
which can be easily checked both at the level of the Feynman parameterizations and MB representations above. We will show in the next section that this is also the case, at the level of each term in the linear combinations that form the solutions that we present in this work.

\section{Solution \label{results}}

We will now extract series representations of both the 9-fold MB integrals shown in Eqs.(\ref{DB}) and (\ref{hex}). This can be done in a systematic way following our
new simple and general method that is based on the construction given in \cite{TZ,Friot:2011ic}, and which will be presented in detail in \cite{ABFG}.
As already mentioned in the introduction, an important advantage of our method is that the derivation of the linear combinations of multiple hypergeometric series that give the different series representations of $N$-fold MB integrals does not require convergence considerations. 
In this sense, our method extends, in a geometrical way, the graphical computational approach of \cite{TZ,Friot:2011ic} dedicated to 2-fold MB integrals. It is straightforward to derive the possible cones as well as their total number from our procedure, and therefore to obtain the different analytic continuations, from the series point of view, that can be extracted for a given $N$-fold MB integral. We have checked for this in many cases that are simpler than the double box and hexagon. However, for the 9-fold MB integrals under study in the present paper, deriving all the cones was much too time consuming. Indeed, applying our method to the hexagon, for example, we have obtained hundreds of different series representations built from linear combinations of 26 series that belong to a set of 2530 possible series (note that this total number of possible series has been also obtained from the Yangian analysis of \cite{Loebbert:2019vcj}). But we know that many more can be derived. It is therefore impossible to list all these possible linear combinations explicitly in one paper and we focus here on the series representations of the double box and hexagon associated to one particular cone.

For the double box case, the result is
\begin{align}
\phi_{3,3}&=\frac{Q_{3,3}\ u_8^{D/2-\ell}}{\Gamma(D/2-f)\Gamma(f)}(D_1+D_{4}+D_{10}+D_{18}+D_{27}+D_{45}+D_{76}+D_{140}+D_{158}+D_{190}\nonumber\\
&+D_{208}+D_{318}+D_{340}+D_{440}+D_{542}+D_{674}+D_{1063}+D_{1091}+D_{1435}+D_{1581}+D_{1646}\nonumber\\
&+D_{2382}+D_{3047}+D_{3068}+D_{3786}+D_{4580}+D_{5}+D_{19}+D_{28}+D_{46}\nonumber\\
&+D_{142}+D_{210}+D_{826}+D_{926}+D_{942}+D_{988}+D_{1094}+D_{1112}+D_{1330}+D_{1449}\nonumber\\
&+D_{1647}+D_{2436}+D_{3069}+D_{3806})
\label{results_I33}
\end{align}
where the $D_i$ are explicitly given in Appendix A. The index $i$ of the $D_i$ corresponds to their ranking in our list of 4834 series that are used to constitute the different series representations of the double box.

The series representation of the hexagon that corresponds to the series shown in Eq. (\ref{results_I33}) (from the differential equation (\ref{diffeq}) point of view), is
\begin{align}
\phi_6&=\frac{Q_6}{\Gamma(D/2-f)\Gamma(f)}(H_1+H_4+H_9+H_{15}+H_{22}+H_{35}+H_{58}+H_{94}+H_{103}+H_{123}\nonumber\\&+H_{133}+H_{199}+H_{210}+H_{270}+H_{331}+H_{409}+H_{637}+H_{653}\nonumber\\&+H_{838}+H_{925}+H_{960}+H_{1375}+H_{1664}+H_{1675}+H_{2062}+H_{2442})
\label{results_I6}
\end{align}
The $H_i$ are also explicitly given, in Appendix B. As for the $D_i$, the index $i$ of the $H_i$ corresponds to their ranking in our list of 2530 series that are used to constitute the different series representations of the hexagon.

It is straightforward to see that, taking into account the overall factor, for each of the first 26 individual terms of Eq.(\ref{results_I33}) that we had reordered to ease this exercise, the differential equation (\ref{diffeq}) is satisfied when compared to the 26 individual terms of the hexagon shown in Eq.(\ref{results_I6}). For the 18 remaining terms of the double box, the differential equation is also satisfied because the derivative of the latter with respect to $u_8$ gives zero.

Finding the convergence region of the series representation presented in Eq.(\ref{results_I33}) (resp.(\ref{results_I6})) from the intersection of the convergence regions of its 44 series (resp. 26 series) is a very hard problem that we have not solved yet. However, our method also allows to derive one particular series, that we call the master series, whose convergence region we conjecture to be included in both the convergence regions of the series representations given in Eqs.(\ref{results_I33}) and (\ref{results_I6}) (the detailed derivation of master series will be discussed in \cite{ABFG}). The latter master series has the following characteristic list \cite{Srivastava}
	\begin{align}
	    &\{n_1+n_2-n_3,n_3-n_2-n_5,n_3+n_4+n_5-n_1-n_6,n_6-n_3,n_1+n_2+n_6-n_3-n_4-n_7, \nonumber\\ &n_6-n_5-n_8,n_7+n_8-n_6,n_3-n_9,n_4+n_5+n_7+n_8-n_6-n_9,n_9,n_9-n_8,n_3,n_5,n_6,n_8,n_9\}
	    \label{charlist}
	\end{align}
	and its arguments are in the order: $\dfrac{w_1w_6}{w_3w_8}$, $\dfrac{w_2w_5}{w_3w_8}$, $\dfrac{w_3w_8}{w_5w_6}$, $\dfrac{w_4w_8}{w_6w_9}$, $\dfrac{w_5w_7}{w_6w_9}$, $\dfrac{w_6w_9}{w_7w_8}$, $\dfrac{w_8}{w_9}$, $\dfrac{w_7}{w_9}$ and $w_9$.

\section{Conclusions}

Series representations of the six point off-shell massless double box with generic propagator powers, and related hexagon, have been presented in Eqs.(\ref{results_I33}) and (\ref{results_I6}). These series representations, which have the form of linear combinations of multiple hypergeometric series of the generalized Horn type, belong to a large set of different series representations that can be systematically derived from our new powerful and simple method of evaluation of multiple MB integrals. All the series representations of this set are analytic continuations of one another. As a check of the results presented in Eqs. (\ref{results_I33}) and (\ref{results_I6}), we have shown that each of their individual terms satisfy the differential equation (\ref{diffeq}). We hope that the solutions of the double box and hexagon that we have obtained from our MB analysis may yield insights for solving the corresponding Yangian constraints.

Our method to derive these series representations, that we will describe in a detailed way in \cite{ABFG}, is a direct extension to the $N$-fold case of the practical computational approach developed in \cite{TZ,Friot:2011ic} for the 2-fold case. It has the great advantage of selecting the different series that constitute a given linear combination forming a series representation of the $N$-fold MB integral under study without the need of a prior knowledge of the convergence regions of the possible involved series. 

Finding the convergence regions of the 9-fold hypergeometric series that constitute the series representations presented in this letter is a difficult open problem, but we would like to add that, for each of the linear combinations that our method can produce, one can also derive a master series whose convergence region, we believe, is common to the convergence regions of all the series that constitute the corresponding linear combination (note that the double box and hexagon series representations in Eqs.(\ref{results_I33}) and (\ref{results_I6}) have the same master series, see Eq.(\ref{charlist})). This can dramatically simplify the convergence analysis in the cases where the convergence region of the master series is equal to the convergence region of the series representation and, in the other cases, this considerably simplifies the numerical checks.

\bigskip

\bigskip

\noindent {\bf Acknowledgements}

\vspace{0.5cm}

We thank Alankar Dutta and Prateek Sharma for technical assistance. S.G. thanks Ulf-G. Meissner for supporting the research through grants.

\newpage

{\bf SUPPLEMENTAL MATERIAL}

\bigskip

\bigskip

{\bf APPENDIX A The double box}

\bigskip

\bigskip

We give here the explicit expressions of the 44 terms that constitute the series representation of the double box in Eq.(\ref{results_I33}). The first 26 terms in the list below correspond, in order, to the set of 26 terms of the hexagon given in Appendix B from the differential equation (\ref{diffeq}) point of view, that they satisfy, individually. Note that several of the $D_i$ are in fact zero because of the occurence of a diverging gamma function in their denominator. However it is necessary to keep them in order to check the differential equation.

The Srivastava-Daoust notation \cite{Srivastava} does not yield shorter expressions. As an example, the $D_4$ series given in Eq.(\ref{D4}) can be written, in the latter notation, as 

\begin{multline}
   D_4 = (w_9)^{-D+e+f+l}
   \frac{\Gamma(c) \Gamma(d) \Gamma(e) \Gamma(D-e-f-l) \Gamma(D/2-e-f)
\Gamma(b-D+e+f+l)}{\Gamma(D-e-f-l)} \\
   \times \Gamma(D-b-c-d-e-l)
   \times F^{7:0;...;0}_{1:0;...;0}
   \left( -w_1,-w_2,-w_3,-\frac{w_4}{w_9},-\frac{w_5}{w_9},-\frac{w_6}{w_9},-\frac{w_7}{w_9},-\frac{w_8}{w_9},-w_9
 \right)
\end{multline}
with
\begin{align*}
    \Omega_1(n_1,...,n_9)
    = \frac{\prod_{j=1}^7 (p_j)_{\theta_j^{(k)} n_k}}
    {\prod_{j=1}^1 (q_j)_{\psi_j^{(k)} n_k}}
\end{align*}
where
\begin{align}
    & p = (c, D-e-f-l, D/2-e-f, b-D+e+f+l, d, D-b-c-d-e-l, e) \nonumber \\
    & q=(D-e-f-l) \nonumber \\
    & \theta_1 = (0,1,0,0,0,1,1,1,0) \nonumber \\
    & \theta_2 = (0,0,0,1,1,1,1,1,-1) \nonumber \\
    & \theta_3 = (0,0,0,1,1,1,0,0,-1) \nonumber \\
    & \theta_4 = (1,0,0,-1,0,-1,-1,0,1) \nonumber \\
    & \theta_5 = (0,0,1,1,1,1,0,0,0) \nonumber \\
    & \theta_6 = (-1,-1,-1,0,-1,-1,0,-1,0) \nonumber \\
    & \theta_7 = (1,1,1,0,0,0,0,0,1) \nonumber \\
    & \psi_1 = (0,0,0,1,1,1,0,0,-1) \nonumber \\
\end{align}
We therefore stick to a more usual notation for the series in the following.

\begin{align}
   D_{1} &= \sum_{n_1=0}^{\infty} ... \sum_{n_9=0}^{\infty}
    \frac{(-w_1)^{n_1}}{n_1!} 
    \frac{(-w_2)^{n_2}}{n_2!}
    \frac{(-w_3)^{n_3}}{n_3!} \frac{(-w_4)^{n_4}}{n_4!}
    \frac{(-w_5)^{n_5}}{n_5!}
    \nonumber \\
    & \times
    \frac{(-w_6)^{n_6}}{n_6!}
    \frac{(-w_7)^{n_7}}{n_7!}
    \frac{(-w_8)^{n_8}}{n_8!}
    \frac{(-w_9)^{n_9}}{n_9!}
    \frac{\Gamma \left(b+n_1+n_5+n_8+n_9\right)}
    {\Gamma \left(-n_7-n_8-n_9\right)}
    \nonumber \\
    & \times 
   \Gamma \left(c+n_2+n_6+n_7+n_8\right)\Gamma \left(D-f-l+n_1+n_2+n_3+n_4+n_5+n_6+n_7+n_8+n_9\right)
        \nonumber \\
    & \times
    \Gamma
   \left(-D/2+l-n_7-n_8-n_9\right) \Gamma\left(-D+e+f+l-n_4-n_5-n_6-n_7-n_8-n_9\right)
        \nonumber \\
    & \times
    \Gamma \left(d+n_3+n_4+n_5+n_6\right)
    \Gamma \left(-b-c-d+D-e-l-n_1-n_2-n_3-n_5-n_6-n_8\right)
\end{align}

\begin{align}
   D_{4} &= (w_9)^{-D+e+f+l} \sum_{n_1=0}^{\infty} ... \sum_{n_9=0}^{\infty}
    \frac{(-w_1)^{n_1}}{n_1!} 
    \frac{(-w_2)^{n_2}}{n_2!}
    \frac{(-w_3)^{n_3}}{n_3!} \frac{(-w_4/w_9)^{n_4}}{n_4!}
    \frac{(-w_5/w_9)^{n_5}}{n_5!}
    \nonumber \\
    & \times
    \frac{(-w_6/w_9)^{n_6}}{n_6!}
    \frac{(-w_7/w_9)^{n_7}}{n_7!}
    \frac{(-w_8/w_9)^{n_8}}{n_8!}
    \frac{(-w_9)^{n_9}}{n_9!}
    \frac{\Gamma \left(c+n_2+n_6+n_7+n_8\right)}
    {\Gamma \left(D-e-f-l+n_4+n_5+n_6-n_9\right)}
    \nonumber \\
    & \times 
    \Gamma \left(D-e-f-l+n_4+n_5+n_6+n_7+n_8-n_9\right)
    \Gamma \left(D/2-e-f+n_4+n_5+n_6-n_9\right) 
        \nonumber \\
    & \times
    \Gamma \left(b-D+e+f+l+n_1-n_4-n_6-n_7+n_9\right)
    \Gamma \left(d+n_3+n_4+n_5+n_6\right)
        \nonumber \\
    & \times
    \Gamma \left(-b-c-d+D-e-l-n_1-n_2-n_3-n_5-n_6-n_8\right)
    \Gamma \left(e+n_1+n_2+n_3+n_9\right)
    \label{D4}
\end{align}

\begin{align}
   D_{10} &= (w_8)^{-b-c-d+D-e-l} \sum_{n_1=0}^{\infty} ... \sum_{n_9=0}^{\infty}
    \frac{(-w_1/w_8)^{n_1}}{n_1!} 
    \frac{(-w_2/w_8)^{n_2}}{n_2!}
    \frac{(-w_3/w_8)^{n_3}}{n_3!} 
    \frac{(-w_4)^{n_4}}{n_4!}
        \nonumber \\
    & \times
    \frac{(-w_5/w_8)^{n_5}}{n_5!}
    \frac{(-w_6/w_8)^{n_6}}{n_6!}
    \frac{(-w_7)^{n_7}}{n_7!}
    \frac{(-w_9)^{n_8}}{n_8!}
    \frac{(-w_8)^{n_9}}{n_9!}
    \Gamma \left(d+n_3+n_4+n_5+n_6\right)
        \nonumber \\
    & \times
    \frac{\Gamma \left(-b-d+D-e-l-n_1-n_3-n_5+n_7+n_9\right)}
    {\Gamma \left(b+c+d-D+e+l+n_1+n_2+n_3+n_5+n_6-n_7-n_8-n_9\right)}
        \nonumber \\
    & \times    
    \Gamma \left(-c-d+D-e-l-n_2-n_3-n_6+n_8+n_9\right)
        \nonumber \\
    & \times   
    \Gamma \left(b+c+d-D+e+l+n_1+n_2+n_3+n_5+n_6-n_9\right)
        \nonumber \\
    & \times   
    \Gamma \left(b+c+d+e+2l+n_1+n_2+n_3+n_5+n_6-n_7-n_8-n_9-3D/2\right)
        \nonumber \\
    & \times \Gamma \left(b+c+d-2 D+2 e+f+2 l+n_1+n_2+n_3-n_4-n_7-n_8-n_9\right)
        \nonumber \\
    & \times 
    \Gamma \left(-b-c-d+2 D-e-f-2 l+n_4+n_7+n_8+n_9\right)
\end{align}

\begin{align}
   D_{18} &= (w_8)^{-b-c-d+D-e-l} (w_9)^{b+c+d-2D+2e+f+2l} \sum_{n_1=0}^{\infty} ... \sum_{n_9=0}^{\infty}
    \frac{(-w_1 w_9/w_8)^{n_1}}{n_1!} 
    \frac{(-w_2 w_9/w_8)^{n_2}}{n_2!}
        \nonumber \\
    & \times
    \frac{(-w_3 w_9/w_8)^{n_3}}{n_3!} 
    \frac{(-w_4/w_9)^{n_4}}{n_4!}
    \frac{(-w_5/w_8)^{n_5}}{n_5!}
    \frac{(-w_6/w_8)^{n_6}}{n_6!}
    \frac{(-w_7/w_9)^{n_7}}{n_7!}
    \frac{(-w_9)^{n_8}}{n_8!}
    \frac{(-w_8/w_9)^{n_9}}{n_9!}
        \nonumber \\
    & \times
    \frac{\Gamma \left(D/2-e-f+n_4+n_5+n_6-n_8\right)}
    {\Gamma \left(D-e-f-l+n_4+n_5+n_6-n_8\right)}
    \Gamma \left(b-D+e+f+l+n_1-n_4-n_6-n_7+n_8\right)
        \nonumber \\
    & \times
    \Gamma \left(-b-d+D-e-l-n_1-n_3-n_5+n_7+n_9\right)
    \Gamma \left(d+n_3+n_4+n_5+n_6\right)
        \nonumber \\
    & \times 
    \Gamma \left(b+c+d-D+e+l+n_1+n_2+n_3+n_5+n_6-n_9\right)
    \Gamma \left(e+n_1+n_2+n_3+n_8\right)    
        \nonumber \\
    & \times
    \Gamma \left(-b-c-d+2D-2 e-f-2 l-n_1-n_2-n_3+n_4+n_7-n_8+n_9\right)
\end{align}

\begin{align}
   D_{27} &= (w_7)^{b-D+e+f+l} (w_9)^{-b} \sum_{n_1=0}^{\infty} ... \sum_{n_9=0}^{\infty}
    \frac{(-w_1 w_7/w_9)^{n_1}}{n_1!} 
    \frac{(-w_2)^{n_2}}{n_2!}
    \frac{(-w_3)^{n_3}}{n_3!} 
    \frac{(-w_4/w_7)^{n_4}}{n_4!}
    \frac{(-w_5/w_9)^{n_5}}{n_5!}    
        \nonumber \\
    & \times
    \frac{(-w_6/w_7)^{n_6}}{n_6!}
    \frac{(-w_8/w_9)^{n_7}}{n_7!}
    \frac{(-w_7/w_9)^{n_8}}{n_8!}
    \frac{(-w_7)^{n_9}}{n_9!}
    \frac{\Gamma \left(D/2-e-f+n_4+n_5+n_6-n_9\right)}
    {\Gamma \left(D-e-f-l+n_4+n_5+n_6-n_9\right)}
        \nonumber \\
    & \times
    \Gamma \left(-b+D-e-f-l-n_1+n_4+n_6-n_8-n_9\right)
    \Gamma \left(b+n_1+n_5+n_7+n_8\right)    
        \nonumber \\
    & \times
    \Gamma \left(-b-c-d+D-e-l-n_1-n_2-n_3-n_5-n_6-n_7\right)
    \Gamma \left(d+n_3+n_4+n_5+n_6\right)    
        \nonumber \\
    & \times
    \Gamma \left(b+c-D+e+f+l+n_1+n_2-n_4+n_7+n_8+n_9\right)
    \Gamma \left(e+n_1+n_2+n_3+n_9\right)    
\end{align}

\begin{align}
   D_{45} &= (w_7)^{b-D+e+f+l} (w_8)^{-b-c-d+D-e-l} (w_9)^{c+d-D+e+l}  \sum_{n_1=0}^{\infty} ... \sum_{n_9=0}^{\infty}
    \frac{(-w_1 w_7/w_8)^{n_1}}{n_1!} 
        \nonumber \\
    & \times
    \frac{(-w_2 w_9/w_8)^{n_2}}{n_2!}
    \frac{(-w_3 w_9/w_8)^{n_3}}{n_3!} 
    \frac{(-w_4/w_7)^{n_4}}{n_4!}
    \frac{(-w_5/w_8)^{n_5}}{n_5!}    
    \frac{(-w_6 w_9/(w_7 w_8))^{n_6}}{n_6!}
        \nonumber \\
    & \times
    \frac{(-w_7/w_9)^{n_7}}{n_7!}
    \frac{(-w_7)^{n_8}}{n_8!}
    \frac{(-w_8/w_9)^{n_9}}{n_9!}
    \frac{\Gamma \left(-d+f-n_3-n_4-n_5-n_6+n_7+n_8+n_9\right)}
    {\Gamma \left(D-e-f-l+n_4+n_5+n_6-n_8\right)}
        \nonumber \\
    & \times
    \Gamma \left(D/2-e-f+n_4+n_5+n_6-n_8\right)
    \Gamma \left(-b+D-e-f-l-n_1+n_4+n_6-n_7-n_8\right)
        \nonumber \\
    & \times
    \Gamma \left(-c-d+D-e-l-n_2-n_3-n_6+n_7+n_9\right)
    \Gamma \left(e+n_1+n_2+n_3+n_8\right)
        \nonumber \\
    & \times    
    \Gamma \left(b+c+d-D+e+l+n_1+n_2+n_3+n_5+n_6-n_9\right)
    \Gamma \left(d+n_3+n_4+n_5+n_6\right)
\end{align}

\begin{align}
   D_{76} &= (w_6)^{-c-d+D-e-l} (w_8)^{-b}  \sum_{n_1=0}^{\infty} ... \sum_{n_9=0}^{\infty}
    \frac{(-w_1/w_8)^{n_1}}{n_1!} 
    \frac{(-w_2/w_6)^{n_2}}{n_2!}  
    \frac{(-w_3/w_6)^{n_3}}{n_3!} 
    \frac{(-w_4)^{n_4}}{n_4!}
        \nonumber \\
    & \times
    \frac{(-w_5/w_8)^{n_5}}{n_5!}    
    \frac{(-w_7)^{n_6}}{n_6!}
    \frac{(-w_6 w_9/w_8)^{n_7}}{n_7!}
    \frac{(-w_6/w_8)^{n_8}}{n_8!}
    \frac{(-w_6)^{n_9}}{n_9!}
    \frac{\Gamma \left(b+n_1+n_5+n_7+n_8\right)}
    {\Gamma \left(b+n_1+n_5-n_6+n_8\right)}    
        \nonumber \\
    & \times
    \Gamma \left(b+l+n_1+n_5-n_6+n_8-D/2\right)
    \Gamma \left(-c+D-e-l-n_2+n_4+n_5+n_7+n_8+n_9\right)
        \nonumber \\
    & \times
    \Gamma \left(-b-d+D-e-l-n_1-n_3-n_5+n_6+n_9\right)
     \nonumber \\
    & \times\Gamma \left(c+d-D+e+l+n_2+n_3-n_7-n_8-n_9\right)
        \nonumber \\
    & \times
    \Gamma \left(b+c+d-2D+2 e+f+2 l+n_1+n_2+n_3-n_4-n_6-n_7-n_9\right)             \nonumber \\
    & \times \Gamma \left(-b-c-d+2D-e-f-2 l+n_4+n_6+n_7+n_9\right)
\end{align}

\begin{align}
   D_{140} &= (w_6)^{-c-d+D-e-l} (w_7)^{b+c+d-2D+2e+f+2l} (w_8)^{-b}  \sum_{n_1=0}^{\infty} ... \sum_{n_9=0}^{\infty}
    \frac{(-w_1 w_7/w_8)^{n_1}}{n_1!} 
    \frac{(-w_2 w_7/w_6)^{n_2}}{n_2!}  
        \nonumber \\
    & \times
    \frac{(-w_3 w_7/w_6)^{n_3}}{n_3!} 
    \frac{(-w_4/w_7)^{n_4}}{n_4!}
    \frac{(-w_5/w_8)^{n_5}}{n_5!}    
    \frac{(-w_6 w_9/(w_7 w_8))^{n_6}}{n_6!}
    \frac{(-w_6/w_8)^{n_7}}{n_7!}
    \frac{(-w_7)^{n_8}}{n_8!}
        \nonumber \\
    & \times
    \frac{(-w_6/w_7)^{n_9}}{n_9!}
    \frac{\Gamma \left(-c+D-e-l-n_2+n_4+n_5+n_6+n_7+n_9\right)
    \Gamma \left(b+n_1+n_5+n_6+n_7\right)}
    {\Gamma \left(-c-d+2D-2 e-f-2l-n_2-n_3+n_4+n_5+n_6+n_7-n_8+n_9\right)}
        \nonumber \\
    & \times
    \Gamma \left(e+n_1+n_2+n_3+n_8\right)
    \Gamma \left(c+d-D+e+l+n_2+n_3-n_6-n_7-n_9\right)
        \nonumber \\
    & \times
    \Gamma \left(c-D+e+f+l+n_2-n_4-n_5-n_6+n_8\right)
        \nonumber \\
    & \times
    \Gamma \left(-c-d+3D/2-2 e-f-l-n_2-n_3+n_4+n_5+n_6+n_7-n_8+n_9\right)
        \nonumber \\
    & \times\Gamma \left(-b-c-d+2D-2 e-f-2 l-n_1-n_2-n_3+n_4+n_6-n_8+n_9\right)
\end{align}

\begin{align}
   D_{158} &= (w_6)^{-d+f} (w_7)^{b+d-D+e+l} (w_8)^{-b-c+D-e-f-l} (w_9)^{c-D+e+f+l}
   \sum_{n_1=0}^{\infty} ... \sum_{n_9=0}^{\infty}
    \frac{(-w_1 w_7/w_8)^{n_1}}{n_1!} 
        \nonumber \\
    & \times
    \frac{(-w_2 w_9/w_8)^{n_2}}{n_2!}
    \frac{(-w_3 w_7/w_6)^{n_3}}{n_3!}
    \frac{(-w_4 w_8/(w_6 w_9))^{n_4}}{n_4!}
    \frac{(-w_5 w_7/(w_6 w_9))^{n_5}}{n_5!}    
    \frac{(-w_6/w_8)^{n_6}}{n_6!}
        \nonumber \\
    & \times
    \frac{(-w_6 w_9/(w_7 w_8))^{n_7}}{n_7!}
    \frac{(-w_6 w_9/w_8)^{n_8}}{n_8!}
    \frac{(-w_6/w_7)^{n_9}}{n_9!}
    \frac{\Gamma \left(-d+D/2-e-n_3+n_6+n_7+n_9\right)}
    {\Gamma \left(-d+D-e-l-n_3+n_6+n_7+n_9\right)}
        \nonumber \\
    & \times
    \Gamma \left(d-f+n_3+n_4+n_5-n_6-n_7-n_8-n_9\right)
    \Gamma \left(-b-d+D-e-l-n_1-n_3-n_5+n_7+n_9\right)
        \nonumber \\
    & \times
    \Gamma \left(-c+D-e-f-l-n_2+n_4+n_5-n_7-n_8\right)
    \Gamma \left(e+n_1+n_2+n_3+n_8\right)    
        \nonumber \\
    & \times    
    \Gamma \left(b+c-D+e+f+l+n_1+n_2-n_4+n_6+n_7+n_8\right)
    \Gamma \left(f+n_6+n_7+n_8+n_9\right)
\end{align}

\begin{align}
   D_{190} &= (w_5)^{-b-d+D-e-l} (w_8)^{-c}
   \sum_{n_1=0}^{\infty} ... \sum_{n_9=0}^{\infty}
    \frac{(-w_1/w_5)^{n_1}}{n_1!}
    \frac{(-w_2/w_8)^{n_2}}{n_2!} 
    \frac{(-w_3/w_5)^{n_3}}{n_3!}
    \frac{(-w_4)^{n_4}}{n_4!}
        \nonumber \\
    & \times
    \frac{(-w_6/w_8)^{n_5}}{n_5!}
    \frac{(-w_5 w_7/w_8)^{n_6}}{n_6!}
    \frac{(-w_9)^{n_7}}{n_7!}
    \frac{(-w_5/w_8)^{n_8}}{n_8!}
    \frac{(-w_5)^{n_9}}{n_9!}
    \frac{\Gamma \left(c+n_2+n_5+n_6+n_8\right)}{\Gamma \left(c+n_2+n_5-n_7+n_8\right)}
        \nonumber \\
    & \times
    \Gamma \left(-b+D-e-l-n_1+n_4+n_5+n_6+n_8+n_9\right)
    \Gamma \left(c+l+n_2+n_5-n_7+n_8-D/2\right)    
        \nonumber \\
    & \times
    \Gamma \left(b+d-D+e+l+n_1+n_3-n_6-n_8-n_9\right)
    \Gamma \left(-c-d+D-e-l-n_2-n_3-n_5+n_7+n_9\right)
        \nonumber \\
    & \times
    \Gamma \left(b+c+d-2D+2 e+f+2 l+n_1+n_2+n_3-n_4-n_6-n_7-n_9\right)        \nonumber \\
    & \times
    \Gamma \left(-b-c-d+2D-e-f-2 l+n_4+n_6+n_7+n_9\right)
\end{align}

\begin{align}
   D_{208} &= (w_5)^{-b-d+D-e-l} (w_8)^{-c} (w_9)^{b+c+d-2D+2e+f+2l} 
   \sum_{n_1=0}^{\infty} ... \sum_{n_9=0}^{\infty}
    \frac{(-w_1 w_9/w_5)^{n_1}}{n_1!}
    \frac{(-w_2 w_9/w_8)^{n_2}}{n_2!} 
        \nonumber \\
    & \times
    \frac{(-w_3 w_9/w_5)^{n_3}}{n_3!}
    \frac{(-w_4/w_9)^{n_4}}{n_4!}
    \frac{(-w_6/w_8)^{n_5}}{n_5!}
    \frac{(-w_5 w_7/(w_8 w_9))^{n_6}}{n_6!}
    \frac{(-w_5/w_8)^{n_7}}{n_7!}
    \frac{(-w_9)^{n_8}}{n_8!}
        \nonumber \\
    & \times
    \frac{(-w_5/w_9)^{n_9}}{n_9!}
    \frac{\Gamma \left(-b-d+3D/2-2e-f-l-n_1-n_3+n_4+n_5+n_6+n_7-n_8+n_9\right)}
    {\Gamma \left(-b-d+2D-2 e-f-2l-n_1-n_3+n_4+n_5+n_6+n_7-n_8+n_9\right)}
        \nonumber \\
    & \times
    \Gamma \left(e+n_1+n_2+n_3+n_8\right)
    \Gamma \left(-b+D-e-l-n_1+n_4+n_5+n_6+n_7+n_9\right)
        \nonumber \\
    & \times    
    \Gamma \left(b+d-D+e+l+n_1+n_3-n_6-n_7-n_9\right)
    \Gamma \left(c+n_2+n_5+n_6+n_7\right)
        \nonumber \\
    & \times 
    \Gamma \left(-b-c-d+2D-2 e-f-2l-n_1-n_2-n_3+n_4+n_6-n_8+n_9\right)         \nonumber \\
    & \times     
    \Gamma \left(b-D+e+f+l+n_1-n_4-n_5-n_6+n_8\right)
\end{align}

\begin{align}
   D_{318} &= (w_5)^{-b-d+D-e-l} (w_6)^{-c-d+D-e-l} (w_8)^{d-D+e+l} 
   \sum_{n_1=0}^{\infty} ... \sum_{n_9=0}^{\infty}
    \frac{(-w_1/w_5)^{n_1}}{n_1!}
    \frac{(-w_2/w_6)^{n_2}}{n_2!}
        \nonumber \\
    & \times
    \frac{(-w_3 w_8/(w_5 w_6))^{n_3}}{n_3!}    
    \frac{(-w_4)^{n_4}}{n_4!}
    \frac{(-w_5 w_7/w_8)^{n_5}}{n_5!}
    \frac{(-w_6 w_9/w_8)^{n_6}}{n_6!}
    \frac{(-w_6/w_8)^{n_7}}{n_7!}
    \frac{(-w_5/w_8)^{n_8}}{n_8!}
        \nonumber \\
    & \times
    \frac{(-w_5 w_6/w_8)^{n_9}}{n_9!}
    \frac{\Gamma \left(-b-c-d+2 D-e-f-2 l+n_4+n_5+n_6+n_9\right)}
    {\Gamma \left(-d+D-e-l-n_3+n_7+n_8+n_9\right)}
        \nonumber \\
    & \times
    \Gamma \left(-d+D-e-l-n_3+n_5+n_6+n_7+n_8+n_9\right)
    \Gamma \left(-d+D/2-e-n_3+n_7+n_8+n_9\right)    
        \nonumber \\
    & \times    
    \Gamma \left(b+d-D+e+l+n_1+n_3-n_5-n_8-n_9\right)
    \Gamma \left(c+d-D+e+l+n_2+n_3-n_6-n_7-n_9\right)        
        \nonumber \\
    & \times
    \Gamma \left(-b-c-d+2D-2 e-2 l-n_1-n_2-n_3+n_4+n_5+n_6+n_7+n_8+2 n_9\right)
        \nonumber \\
    & \times    
    \Gamma \left(b+c+d-2D+2 e+f+2 l+n_1+n_2+n_3-n_4-n_5-n_6-n_9\right)
\end{align}

\begin{align}
   D_{340} &= (w_5)^{-b-d+D-e-l} (w_6)^{b-D+e+f+l} (w_8)^{-b-c+D-e-f-l} (w_9)^{b+c+d-2D+2e+f+2l} 
          \nonumber \\
    & \times
   \sum_{n_1=0}^{\infty} ... \sum_{n_9=0}^{\infty}
    \frac{(-w_1 w_6 w_9/(w_5 w_8))^{n_1}}{n_1!}
    \frac{(-w_2 w_9/w_8)^{n_2}}{n_2!}
    \frac{(-w_3 w_9/w_5)^{n_3}}{n_3!}    
    \frac{(-w_4 w_8/(w_6 w_9))^{n_4}}{n_4!}
        \nonumber \\
    & \times
    \frac{(-w_5 w_7/(w_6 w_9))^{n_5}}{n_5!}     
    \frac{(-w_6/w_8)^{n_6}}{n_6!}
    \frac{(-w_5/w_8)^{n_7}}{n_7!}
    \frac{(-w_6 w_9/w_8)^{n_8}}{n_8!}
    \frac{(-w_5/w_9)^{n_9}}{n_9!}    
        \nonumber \\
    & \times
    \frac{\Gamma \left(-d+D/2-e-n_3+n_6+n_7+n_9\right)}
    {\Gamma \left(-d+D-e-l-n_3+n_6+n_7+n_9\right)}
    \Gamma \left(-b+D-e-f-l-n_1+n_4+n_5-n_6-n_8\right)
        \nonumber \\
    & \times    
    \Gamma \left(b+d-D+e+l+n_1+n_3-n_5-n_7-n_9\right)
    \Gamma \left(f+n_6+n_7+n_8+n_9\right)    
        \nonumber \\
    & \times
    \Gamma \left(b+c-D+e+f+l+n_1+n_2-n_4+n_6+n_7+n_8\right)
    \Gamma \left(e+n_1+n_2+n_3+n_8\right) 
        \nonumber \\
    & \times     
    \Gamma \left(-b-c-d+2D-2 e-f-2 l-n_1-n_2-n_3+n_4+n_5-n_8+n_9\right)
\end{align}

\begin{align}
   D_{440} &= (w_4)^{b+c-D+e+f+l} (w_7)^{-c} (w_9)^{-b}
   \nonumber \\
    & \times
   \sum_{n_1=0}^{\infty} ... \sum_{n_9=0}^{\infty}
    \frac{(-w_1 w_4/w_9)^{n_1}}{n_1!}
    \frac{(-w_2 w_4/w_7)^{n_2}}{n_2!}
    \frac{(-w_3)^{n_3}}{n_3!}    
    \frac{(-w_5/w_9)^{n_4}}{n_4!}
    \frac{(-w_6/w_7)^{n_5}}{n_5!}     
        \nonumber \\
    & \times
    \frac{(-w_4 w_8/(w_7 w_9))^{n_6}}{n_6!}       
    \frac{(-w_4/w_9)^{n_7}}{n_7!}
    \frac{(-w_4/w_7)^{n_8}}{n_8!}
    \frac{(-w_4)^{n_9}}{n_9!}
        \nonumber \\
    & \times
    \frac{\Gamma \left(b+c+l+n_1+n_2+n_4+n_5+n_6+n_7+n_8-D/2\right)}{\Gamma \left(b+c+n_1+n_2+n_4+n_5+n_6+n_7+n_8\right)}
    \Gamma \left(b+n_1+n_4+n_6+n_7\right)   
        \nonumber \\
    & \times        
    \Gamma \left(-b-c-d+D-e-l-n_1-n_2-n_3-n_4-n_5-n_6\right)
    \Gamma \left(c+n_2+n_5+n_6+n_8\right)    
        \nonumber \\
    & \times        
    \Gamma \left(-b-c+D-e-f-l-n_1-n_2-n_6-n_7-n_8-n_9\right)    
    \Gamma \left(e+n_1+n_2+n_3+n_9\right)
        \nonumber \\
    & \times        
    \Gamma \left(b+c+d-D+e+f+l+n_1+n_2+n_3+n_4+n_5+n_6+n_7+n_8+n_9\right)
\end{align}

\begin{align}
   D_{542} &= (w_4)^{b+c-D+e+f+l} (w_6)^{-b-c-d+D-e-l} (w_7)^{b+d-D+e+l} (w_9)^{-b}
    \sum_{n_1=0}^{\infty} ... \sum_{n_9=0}^{\infty}
    \frac{(-w_1 w_4 w_7/(w_6 w_9))^{n_1}}{n_1!}
       \nonumber \\
    & \times
    \frac{(-w_2 w_4/w_6)^{n_2}}{n_2!}
    \frac{(-w_3 w_7/w_6)^{n_3}}{n_3!}    
    \frac{(-w_5 w_7/(w_6 w_9))^{n_4}}{n_4!}
    \frac{(-w_4 w_8/(w_6 w_9))^{n_5}}{n_5!}
    \frac{(-w_4/w_9)^{n_6}}{n_6!}
        \nonumber \\
    & \times
    \frac{(-w_4/w_7)^{n_7}}{n_7!}
    \frac{(-w_4)^{n_8}}{n_8!}
    \frac{(-w_6/w_7)^{n_9}}{n_9!}
    \frac{\Gamma \left(-d+D/2-e-n_3+n_6+n_7+n_9\right)}
    {\Gamma \left(-d+D-e-l-n_3+n_6+n_7+n_9\right)}
        \nonumber \\
    & \times
    \Gamma \left(-b-d+D-e-l-n_1-n_3-n_4+n_7+n_9\right)    
    \Gamma \left(e+n_1+n_2+n_3+n_8\right)    
        \nonumber \\
    & \times
    \Gamma \left(b+c+d-D+e+l+n_1+n_2+n_3+n_4+n_5-n_9\right)
    \Gamma \left(b+n_1+n_4+n_5+n_6\right)    
        \nonumber \\
    & \times
    \Gamma \left(-b-c+D-e-f-l-n_1-n_2-n_5-n_6-n_7-n_8\right)
    \Gamma \left(f+n_6+n_7+n_8+n_9\right)
\end{align}

\begin{align}
   D_{674} &= (w_4)^{b+c-D+e+f+l} (w_5)^{-b-d+D-e-l} (w_6)^{-c} (w_9)^{d-D+e+l}
    \sum_{n_1=0}^{\infty} ... \sum_{n_9=0}^{\infty}
    \frac{(-w_1 w_4/w_5)^{n_1}}{n_1!}
       \nonumber \\
    & \times
    \frac{(-w_2 w_4/w_6)^{n_2}}{n_2!}
    \frac{(-w_3 w_9/w_5)^{n_3}}{n_3!}    
    \frac{(-w_5 w_7/(w_6 w_9))^{n_4}}{n_4!}
    \frac{(-w_4 w_8/(w_6 w_9))^{n_5}}{n_5!}
    \frac{(-w_4/w_9)^{n_6}}{n_6!}
        \nonumber \\
    & \times
    \frac{(-w_4 w_5/(w_6 w_9))^{n_7}}{n_7!}
    \frac{(-w_4)^{n_8}}{n_8!}
    \frac{(-w_5/w_9)^{n_9}}{n_9!}
    \frac{\Gamma \left(-d+D/2-e-n_3+n_6+n_7+n_9\right)}
    {\Gamma \left(-d+D-e-l-n_3+n_6+n_7+n_9\right)}
        \nonumber \\
    & \times
    \Gamma \left(-d+D-e-l-n_3+n_4+n_5+n_6+n_7+n_9\right)
    \Gamma \left(c+n_2+n_4+n_5+n_7\right)
        \nonumber \\
    & \times    
    \Gamma \left(b+d-D+e+l+n_1+n_3-n_4-n_7-n_9\right)
    \Gamma \left(f+n_6+n_7+n_8+n_9\right)    
        \nonumber \\
    & \times    
    \Gamma \left(-b-c+D-e-f-l-n_1-n_2-n_5-n_6-n_7-n_8\right)
    \Gamma \left(e+n_1+n_2+n_3+n_8\right)    
\end{align}

\begin{align}
   D_{1063} &= (w_3)^{-d+D-e-l} (w_5)^{-b} (w_6)^{-c}
    \sum_{n_1=0}^{\infty} ... \sum_{n_9=0}^{\infty}
    \frac{(-w_1/w_5)^{n_1}}{n_1!}
    \frac{(-w_2/w_6)^{n_2}}{n_2!}
    \frac{(-w_4)^{n_3}}{n_3!}
    \nonumber \\
        & \times
    \frac{(-w_3 w_7/w_6)^{n_4}}{n_4!}    
    \frac{(-w_3 w_8/(w_5 w_6))^{n_5}}{n_5!}
    \frac{(-w_3 w_9/w_5)^{n_6}}{n_6!}
    \frac{(-w_3/w_5)^{n_7}}{n_7!}
    \frac{(-w_3/w_6)^{n_8}}{n_8!}
    \frac{(-w_3)^{n_9}}{n_9!}
        \nonumber \\
    & \times
    \frac{\Gamma \left(b+n_1+n_5+n_6+n_7\right)}
    {\Gamma \left(-n_4-n_5-n_6\right)}
    \Gamma \left(d-D+e+l-n_4-n_5-n_6-n_7-n_8-n_9\right)
        \nonumber \\
    & \times      
    \Gamma \left(b+c-D+e+f+l+n_1+n_2-n_3+n_5+n_7+n_8\right)
    \Gamma \left(-D/2+l-n_4-n_5-n_6\right)
        \nonumber \\
    & \times     
    \Gamma \left(-b-c+D-e-l-n_1-n_2+n_3-n_5+n_9\right)
    \Gamma \left(c+n_2+n_4+n_5+n_8\right)    
        \nonumber \\
    & \times     
    \Gamma \left(-b-c-d+2 D-e-f-2 l+n_3+n_4+n_6+n_9\right)
\end{align}

\begin{align}
   D_{1091} &= (w_3)^{-b-c-d+2D-2e-2l} (w_5)^{c-D+e+l} (w_6)^{b-D+e+l} (w_8)^{-b-c+D-e-l}
    \sum_{n_1=0}^{\infty} ... \sum_{n_9=0}^{\infty}
    \nonumber \\
        & \times    
    \frac{(-w_1 w_6/(w_3 w_8))^{n_1}}{n_1!}\frac{(-w_2 w_5/(w_3 w_8))^{n_2}}{n_2!}
    \frac{(-w_3 w_4 w_8/(w_5 w_6))^{n_3}}{n_3!}
    \frac{(-w_3 w_7/w_6)^{n_4}}{n_4!}    
    \nonumber \\
        & \times
    \frac{(-w_3 w_9/w_5)^{n_5}}{n_5!}
    \frac{(-w_3/w_5)^{n_6}}{n_6!}
    \frac{(-w_3/w_6)^{n_7}}{n_7!}    
    \frac{(-w_3 w_8/(w_5 w_6))^{n_8}}{n_8!}
    \frac{(-w_3^2 w_8/(w_5 w_6))^{n_9}}{n_9!}
        \nonumber \\
    & \times
    \frac
    {\Gamma \left(b+c+e+2 l+n_1+n_2-n_3-n_4-n_5-n_8-n_9-3D/2\right)}
    {\Gamma \left(b+c-D+e+l+n_1+n_2-n_3-n_4-n_5-n_8-n_9\right)}
        \nonumber \\
    & \times
    \Gamma \left(f+n_6+n_7+n_8+n_9\right)
    \Gamma \left(-b+D-e-l-n_1+n_3+n_4+n_7+n_8+n_9\right)
        \nonumber \\
    & \times    
    \Gamma \left(-c+D-e-l-n_2+n_3+n_5+n_6+n_8+n_9\right)
        \nonumber \\
    & \times    
    \Gamma \left(b+c-D+e+l+n_1+n_2-n_3-n_8-n_9\right) 
        \nonumber \\
    & \times    
    \Gamma \left(b+c+d-2D+2 e+2 l+n_1+n_2-n_3-n_4-n_5-n_6-n_7-n_8-2 n_9\right)
        \nonumber \\
    & \times    
    \Gamma \left(-b-c-d+2D-e-f-2 l+n_3+n_4+n_5+n_9\right)
\end{align}

\begin{align}
   D_{1435} &= (w_3)^{-d+D-e-l} (w_4)^{b+c-D+e+f+l} (w_5)^{-b} (w_6)^{-c}
    \sum_{n_1=0}^{\infty} ... \sum_{n_9=0}^{\infty}    
    \frac{(-w_1 w_4/w_5)^{n_1}}{n_1!}
    \frac{(-w_2 w_4/w_6)^{n_2}}{n_2!}
       \nonumber \\
        & \times
    \frac{(-w_3 w_7/w_6)^{n_3}}{n_3!}
    \frac{(-w_3 w_4 w_8/(w_5 w_6))^{n_4}}{n_4!}    
    \frac{(-w_3 w_9/w_5)^{n_5}}{n_5!}
    \frac{(-w_3 w_4/w_5)^{n_6}}{n_6!}
    \frac{(-w_3 w_4/w_6)^{n_7}}{n_7!}
        \nonumber \\
    & \times
    \frac{(-w_4)^{n_8}}{n_8!}
    \frac{(-w_3)^{n_9}}{n_9!}    
    \frac{\Gamma \left(b+n_1+n_4+n_5+n_6\right)}{\Gamma \left(-n_3-n_4-n_5\right)}
    \Gamma \left(d-D+e+l-n_3-n_4-n_5-n_6-n_7-n_9\right)
        \nonumber \\
    & \times    
    \Gamma \left(-D/2+l-n_3-n_4-n_5\right)
    \Gamma \left(-b-c+D-e-f-l-n_1-n_2-n_4-n_6-n_7-n_8\right)
        \nonumber \\
    & \times
        \Gamma \left(-d+D-l+n_1+n_2+n_3+n_4+n_5+n_6+n_7+n_8+n_9\right)
\end{align}

\begin{align}
   D_{1581} &= (w_2)^{-c+D-e-l} (w_6)^{-d} (w_8)^{-b}
    \sum_{n_1=0}^{\infty} ... \sum_{n_9=0}^{\infty}    
    \frac{(-w_1/w_8 )^{n_1}}{n_1!}
       \nonumber \\
        & \times
    \frac{(-w_3/w_6)^{n_2}}{n_2!}    
    \frac{(-w_2 w_4/w_6)^{n_3}}{n_3!}
    \frac{(-w_2 w_5/(w_6 w_8))^{n_4}}{n_4!}    
    \frac{(-w_7)^{n_5}}{n_5!}
    \frac{(-w_2 w_9/w_8)^{n_6}}{n_6!}
        \nonumber \\
    & \times
    \frac{(-w_2/w_8)^{n_7}}{n_7!}    
    \frac{(-w_2/w_6)^{n_8}}{n_8!}
    \frac{(-w_2)^{n_9}}{n_9!}    
    \frac{\Gamma \left(b-D/2+l+n_1+n_4-n_5+n_7\right)}{\Gamma \left(b+n_1+n_4-n_5+n_7\right)}
       \nonumber \\
    & \times
    \Gamma \left(d+n_2+n_3+n_4+n_8\right) \Gamma \left(b+d-D+e+f+l+n_1+n_2+n_4-n_5+n_7+n_8\right)
        \nonumber \\
    & \times
    \Gamma \left(c-D+e+l-n_3-n_4-n_6-n_7-n_8-n_9\right) \Gamma \left(b+n_1+n_4+n_6+n_7\right)
        \nonumber \\
    & \times
    \Gamma \left(-b-c-d+2 D-e-f-2 l+n_3+n_5+n_6+n_9\right) \nonumber \\
    & \times
     \Gamma \left(-b-d+D-e-l-n_1-n_2-n_4+n_5+n_9\right)
\end{align}

\begin{align}
   D_{1646} &= (w_2)^{-c+D-e-l} (w_6)^{-d} (w_7)^{b+d-D+e+f+l} (w_8)^{-b}
    \sum_{n_1=0}^{\infty} ... \sum_{n_9=0}^{\infty}    
    \frac{(-w_1 w_7/w_8 )^{n_1}}{n_1!}
       \nonumber \\
        & \times
    \frac{(-w_3 w_7/w_6)^{n_2}}{n_2!}    
    \frac{(-w_2 w_4/w_6)^{n_3}}{n_3!}
    \frac{(-w_2 w_5 w_7/(w_6 w_8))^{n_4}}{n_4!}    
    \frac{(-w_2 w_9/w_8)^{n_5}}{n_5!}
    \frac{(-w_2 w_7/w_8)^{n_6}}{n_6!}
        \nonumber \\
    & \times
    \frac{(-w_2 w_7/w_6)^{n_7}}{n_7!}    
    \frac{(-w_7)^{n_8}}{n_8!}
    \frac{(-w_2)^{n_9}}{n_9!}    
    \frac{\Gamma \left(b+n_1+n_4+n_5+n_6\right)}{\Gamma \left(-d+D-e-f-l-n_2-n_7-n_8\right)}
        \nonumber \\
    & \times
    \Gamma \left(d+n_2+n_3+n_4+n_7\right)\Gamma \left(-d+D/2-e-f-n_2-n_7-n_8\right) 
        \nonumber \\
    & \times
    \Gamma \left(f+n_6+n_7+n_8+n_9\right) \Gamma \left(c-D+e+l-n_3-n_4-n_5-n_6-n_7-n_9\right)
        \nonumber \\
    & \times
    \Gamma \left(-c+D-l+n_1+n_2+n_3+n_4+n_5+n_6+n_7+n_8+n_9\right)  \nonumber \\
    & \times \Gamma \left(-b-d+D-e-f-l-n_1-n_2-n_4-n_6-n_7-n_8\right)
\end{align}

\begin{align}
   D_{2382} &= (w_2)^{-c+D-e-l} (w_3)^{-b-d+D-e-l} (w_6)^{b-D+e+l}
   (w_8)^{-b}
    \sum_{n_1=0}^{\infty} ... \sum_{n_9=0}^{\infty}    
    \frac{(-w_1 w_6/(w_3 w_8) )^{n_1}}{n_1!}
       \nonumber \\
        & \times
    \frac{(-w_2 w_4/w_6)^{n_2}}{n_2!}    
    \frac{(-w_2 w_5/(w_3 w_8))^{n_3}}{n_3!}
    \frac{(-w_3 w_7/(w_6))^{n_4}}{n_4!}    
    \frac{(-w_2 w_9/w_8)^{n_5}}{n_5!}
    \frac{(-w_2 /w_8)^{n_6}}{n_6!}
        \nonumber \\
    & \times
    \frac{(-w_3/w_6)^{n_7}}{n_7!}    
    \frac{(-w_2/w_6)^{n_8}}{n_8!}
    \frac{(-w_2 w_3 / w_6)^{n_9}}{n_9!}    
    \frac{\Gamma \left(b+n_1+n_3+n_5+n_6\right)}{\Gamma \left(b+n_1+n_3-n_4+n_6\right)}
        \nonumber \\
    & \times
   \Gamma \left(f+n_6+n_7+n_8+n_9\right)
    \Gamma
   \left(-b+D-e-l-n_1+n_2+n_4+n_7+n_8+n_9\right)
        \nonumber \\
    & \times
    \Gamma
   \left(b+d-D+e+l+n_1+n_3-n_4-n_7-n_9\right) \Gamma \left(b-D/2+l+n_1+n_3-n_4+n_6\right)
        \nonumber \\
    & \times
   \Gamma \left(-b-c-d+2 D-e-f-2 l+n_2+n_4+n_5+n_9\right) \nonumber \\
    & \times \Gamma \left(c-D+e+l-n_2-n_3-n_5-n_6-n_8-n_9\right)
\end{align}

 \begin{align}
   D_{3047} &= (w_1)^{-b+D-e-l} (w_5)^{-d} (w_8)^{-c}
    \sum_{n_1=0}^{\infty} ... \sum_{n_9=0}^{\infty}    
    \frac{(-w_2/ w_8 )^{n_1}}{n_1!}
       \nonumber \\
        & \times
    \frac{(-w_3 /w_5)^{n_2}}{n_2!}    
    \frac{(-w_1 w_4 /w_5)^{n_3}}{n_3!}
    \frac{(-w_1 w_6/(w_5 w_8))^{n_4}}{n_4!}    
    \frac{(-w_1 w_7/w_8)^{n_5}}{n_5!}
    \frac{(-w_9)^{n_6}}{n_6!}
        \nonumber \\
    & \times
    \frac{(-w_1 /w_8)^{n_7}}{n_7!}    
    \frac{(-w_1 /w_5)^{n_8}}{n_8!}
    \frac{(-w_1)^{n_9}}{n_9!}    
    \frac{\Gamma \left(c+n_1+n_4+n_5+n_7\right)}{\Gamma \left(c+n_1+n_4-n_6+n_7\right)}
   \Gamma \left(d+n_2+n_3+n_4+n_8\right)
        \nonumber \\
    & \times
   \Gamma \left(c-D/2+l+n_1+n_4-n_6+n_7\right) \Gamma
   \left(b-D+e+l-n_3-n_4-n_5-n_7-n_8-n_9\right)
        \nonumber \\
    & \times
    \Gamma \left(-c-d+D-e-l-n_1-n_2-n_4+n_6+n_9\right) 
        \nonumber \\
    & \times
   \Gamma
   \left(c+d-D+e+f+l+n_1+n_2+n_4-n_6+n_7+n_8\right) \nonumber \\
    & \times \Gamma \left(-b-c-d+2 D-e-f-2 l+n_3+n_5+n_6+n_9\right)
\end{align}
 
 \begin{align}
   D_{3068} &= (w_1)^{-b+D-e-l} (w_5)^{-d} (w_8)^{-c} (w_9)^{c+d-D+e+f+l}
    \sum_{n_1=0}^{\infty} ... \sum_{n_9=0}^{\infty}    
    \frac{(-w_2 w_9/ w_8 )^{n_1}}{n_1!}
       \nonumber \\
        & \times
    \frac{(-w_3 w_9/w_5)^{n_2}}{n_2!}    
    \frac{(-w_1 w_4 /w_5)^{n_3}}{n_3!}
    \frac{(-w_1 w_6 w_9/(w_5 w_8))^{n_4}}{n_4!}    
    \frac{(-w_1 w_7/w_8)^{n_5}}{n_5!}
    \frac{(-w_1 w_9/w_8)^{n_6}}{n_6!}
        \nonumber \\
    & \times
    \frac{(-w_1 w_9/w_5)^{n_7}}{n_7!}    
    \frac{(-w_9)^{n_8}}{n_8!}
    \frac{(-w_1)^{n_9}}{n_9!}    
    \frac{\Gamma \left(c+n_1+n_4+n_5+n_6\right)}{\Gamma \left(-d+D-e-f-l-n_2-n_7-n_8\right)}
        \nonumber \\
    & \times
   \Gamma \left(f+n_6+n_7+n_8+n_9\right) \Gamma
   \left(-b+D-l+n_1+n_2+n_3+n_4+n_5+n_6+n_7+n_8+n_9\right)
        \nonumber \\
    & \times
   \Gamma \left(b-D+e+l-n_3-n_4-n_5-n_6-n_7-n_9\right) \Gamma
   \left(-d+D/2-e-f-n_2-n_7-n_8\right)
        \nonumber \\
    & \times
   \Gamma \left(-c-d+D-e-f-l-n_1-n_2-n_4-n_6-n_7-n_8\right) \Gamma \left(d+n_2+n_3+n_4+n_7\right)
\end{align}

\begin{align}
   D_{3786} &= (w_1)^{-b+D-e-l} (w_3)^{-c-d+D-e-l}(w_5)^{c-D+e+l} (w_8)^{-c}
    \sum_{n_1=0}^{\infty} ... \sum_{n_9=0}^{\infty}    
    \frac{(-w_2 w_5/ (w_3 w_8) )^{n_1}}{n_1!}
       \nonumber \\
        & \times
    \frac{(-w_1 w_4/w_5)^{n_2}}{n_2!}    
    \frac{(-w_1 w_6 /(w_3 w_8))^{n_3}}{n_3!}
    \frac{(-w_1 w_7/w_8)^{n_4}}{n_4!}    
    \frac{(-w_3 w_9/w_5)^{n_5}}{n_5!}
    \frac{(-w_3 /w_5)^{n_6}}{n_6!}
        \nonumber \\
    & \times
    \frac{(-w_1 /w_8)^{n_7}}{n_7!}    
    \frac{(-w_1 /w_5)^{n_8}}{n_8!}
    \frac{(-w_1 w_3 /w_5)^{n_9}}{n_9!}    
    \frac{\Gamma \left(c+n_1+n_3+n_4+n_7\right) }{\Gamma \left(c+n_1+n_3-n_5+n_7\right)}
        \nonumber \\
    & \times
   \Gamma \left(c-D/2+l+n_1+n_3-n_5+n_7\right) \Gamma
   \left(b-D+e+l-n_2-n_3-n_4-n_7-n_8-n_9\right)
        \nonumber \\
    & \times
   \Gamma \left(-c+D-e-l-n_1+n_2+n_5+n_6+n_8+n_9\right)  \Gamma \left(f+n_6+n_7+n_8+n_9\right)
        \nonumber \\
    & \times
  \Gamma \left(-b-c-d+2 D-e-f-2 l+n_2+n_4+n_5+n_9\right)  \nonumber \\
    & \times \Gamma
   \left(c+d-D+e+l+n_1+n_3-n_5-n_6-n_9\right)
\end{align}

\begin{align}
   D_{4580} &= (w_1)^{-b+D-e-l} (w_2)^{-c+D-e-l} (w_3)^{-d} (w_8)^{-D+e+l}
    \sum_{n_1=0}^{\infty} ... \sum_{n_9=0}^{\infty}
    \frac{(-w_1 w_2 w_4/ (w_3 w_8) )^{n_1}}{n_1!}
       \nonumber \\
        & \times
    \frac{(-w_2 w_5/(w_3 w_8))^{n_2}}{n_2!}    
    \frac{(-w_1 w_6 /(w_3 w_8))^{n_3}}{n_3!}
    \frac{(-w_1 w_7/w_8)^{n_4}}{n_4!}    
    \frac{(-w_2 w_9/w_8)^{n_5}}{n_5!}
    \frac{(-w_2 /w_8)^{n_6}}{n_6!}
        \nonumber \\
    & \times
    \frac{(-w_1 /w_8)^{n_7}}{n_7!}    
    \frac{(-w_1 w_2/(w_3 w_8))^{n_8}}{n_8!}
    \frac{(-w_1 w_2/w_8)^{n_9}}{n_9!}   
    \nonumber \\
    & \times
    \frac{\Gamma \left(d+n_1+n_2+n_3+n_8\right) \Gamma \left(D/2-e+n_1+n_2+n_3+n_6+n_7+n_8+n_9\right) }{\Gamma \left(D-e-l+n_1+n_2+n_3+n_6+n_7+n_8+n_9\right)}
        \nonumber \\
    & \times
    \Gamma
   \left(D-e-l+n_1+n_2+n_3+n_4+n_5+n_6+n_7+n_8+n_9\right)
        \nonumber \\
    & \times
   \Gamma \left(b-D+e+l-n_1-n_3-n_4-n_7-n_8-n_9\right) 
        \nonumber \\
    & \times
  \Gamma \left(f+n_6+n_7+n_8+n_9\right) \Gamma \left(-b-c-d+2 D-e-f-2 l+n_1+n_4+n_5+n_9\right) \nonumber \\
    & \times \Gamma
   \left(c-D+e+l-n_1-n_2-n_5-n_6-n_8-n_9\right)
\end{align}

\begin{align}
   D_{5} &= (w_9)^{-D/2+l} \sum_{n_1=0}^{\infty} ... \sum_{n_9=0}^{\infty}
    \frac{(-w_1)^{n_1}}{n_1!} 
    \frac{(-w_2)^{n_2}}{n_2!}
    \frac{(-w_3)^{n_3}}{n_3!} 
    \frac{(-w_4)^{n_4}}{n_4!}
    \frac{(-w_5)^{n_5}}{n_5!}
    \frac{(-w_6)^{n_6}}{n_6!}
    \nonumber \\
    & \times
    \frac{(-w_7/w_9)^{n_7}}{n_7!}
    \frac{(-w_8/w_9)^{n_8}}{n_8!}
    \frac{(-w_9)^{n_9}}{n_9!}
    \frac{\Gamma \left(-D/2+e+f-n_4-n_5-n_6-n_9\right)}
    {\Gamma \left(D/2-l-n_9\right)}
    \nonumber \\
    & \times  
    \Gamma \left(-b-c-d+D-e-l-n_1-n_2-n_3-n_5-n_6-n_8\right)
    \Gamma \left(d+n_3+n_4+n_5+n_6\right)
        \nonumber \\
    & \times 
    \Gamma \left(D/2-f+n_1+n_2+n_3+n_4+n_5+n_6+n_9\right)
    \Gamma \left(c+n_2+n_6+n_7+n_8\right)
        \nonumber \\
    & \times 
    \Gamma \left(D/2-l+n_7+n_8-n_9\right)
    \Gamma \left(b+l+n_1+n_5-n_7+n_9-D/2\right)
\end{align}
\begin{align}
   D_{19} &= (w_8)^{-b-c-d+D-e-l} (w_9)^{b+c+d-3D/2+e+2l} \sum_{n_1=0}^{\infty} ... \sum_{n_9=0}^{\infty}
    \frac{(-w_1 w_9/w_8)^{n_1}}{n_1!} 
    \frac{(-w_2 w_9/w_8)^{n_2}}{n_2!}
        \nonumber \\
    & \times
    \frac{(-w_3 w_9/w_8)^{n_3}}{n_3!} 
    \frac{(-w_4)^{n_4}}{n_4!}
    \frac{(-w_5 w_9/w_8)^{n_5}}{n_5!}
    \frac{(-w_6 w_9/w_8)^{n_6}}{n_6!}
    \frac{(-w_7/w_9)^{n_7}}{n_7!}
    \frac{(-w_8/w_9)^{n_8}}{n_8!}
    \frac{(-w_9)^{n_9}}{n_9!}
        \nonumber \\
    & \times \frac{ \Gamma \left(-b-d+D-e-l-n_1-n_3-n_5+n_7+n_8\right)}
    {\Gamma \left(D/2-l-n_9\right)}
    \Gamma \left(b+l+n_1+n_5-n_7+n_9-D/2\right)
        \nonumber \\
    & \times
    \Gamma \left(D/2-f+n_1+n_2+n_3+n_4+n_5+n_6+n_9\right)
    \Gamma \left(-D/2+e+f-n_4-n_5-n_6-n_9\right)
        \nonumber \\
    & \times 
    \Gamma \left(b+c+d-D+e+l+n_1+n_2+n_3+n_5+n_6-n_8\right)
    \Gamma \left(d+n_3+n_4+n_5+n_6\right)
        \nonumber \\
    & \times 
    \Gamma \left(-b-c-d+3D/2-e-2 l-n_1-n_2-n_3-n_5-n_6+n_7+n_8-n_9\right)
\end{align}
\begin{align}
   D_{28} &= (w_7)^{b-D/2+l} (w_9)^{-b} \sum_{n_1=0}^{\infty} ... \sum_{n_9=0}^{\infty}
    \frac{(-w_1 w_7/w_9)^{n_1}}{n_1!} 
    \frac{(-w_2)^{n_2}}{n_2!}
    \frac{(-w_3)^{n_3}}{n_3!} 
    \frac{(-w_4)^{n_4}}{n_4!}
    \frac{(-w_5 w_7/w_9)^{n_5}}{n_5!}    
        \nonumber \\
    & \times
    \frac{(-w_6)^{n_6}}{n_6!}
    \frac{(-w_8/w_9)^{n_7}}{n_7!}
    \frac{(-w_7/w_9)^{n_8}}{n_8!}
    \frac{(-w_7)^{n_9}}{n_9!}
    \frac{    \Gamma \left(D/2-f+n_1+n_2+n_3+n_4+n_5+n_6+n_9\right)}
    {\Gamma \left(D/2-l-n_9\right)}
        \nonumber \\
    & \times
    \Gamma \left(b+c+l+n_1+n_2+n_5+n_6+n_7+n_8+n_9-D/2\right)
    \Gamma \left(d+n_3+n_4+n_5+n_6\right)
        \nonumber \\
    & \times
    \Gamma \left(-D/2+e+f-n_4-n_5-n_6-n_9\right)
    \Gamma \left(-b+D/2-l-n_1-n_5-n_8-n_9\right)
        \nonumber \\
    & \times
    \Gamma \left(-b-c-d+D-e-l-n_1-n_2-n_3-n_5-n_6-n_7\right)
        \Gamma \left(b+n_1+n_5+n_7+n_8\right)
\end{align}
\begin{align}
   D_{46} &= (w_7)^{b-D/2+l} (w_8)^{-b-c-d+D-e-l} (w_9)^{c+d-D+e+l}  \sum_{n_1=0}^{\infty} ... \sum_{n_9=0}^{\infty}
    \frac{(-w_1 w_7/w_8)^{n_1}}{n_1!} 
    \frac{(-w_2 w_9/w_8)^{n_2}}{n_2!}    
        \nonumber \\
    & \times
    \frac{(-w_3 w_9/w_8)^{n_3}}{n_3!} 
    \frac{(-w_4)^{n_4}}{n_4!}
    \frac{(-w_5 w_7/w_8)^{n_5}}{n_5!}    
    \frac{(-w_6 w_9/w_8)^{n_6}}{n_6!}
    \frac{(-w_7/w_9)^{n_7}}{n_7!}
    \frac{(-w_8/w_9)^{n_8}}{n_8!}
    \frac{(-w_7)^{n_9}}{n_9!}
        \nonumber \\
    & \times
    \frac{\Gamma \left(d+n_3+n_4+n_5+n_6\right)}{\Gamma \left(D/2-l-n_9\right)}
    \Gamma \left(b+c+d-D+e+l+n_1+n_2+n_3+n_5+n_6-n_8\right)
        \nonumber \\
    & \times
    \Gamma \left(-b+D/2-l-n_1-n_5-n_7-n_9\right)
    \Gamma \left(-c-d+D-e-l-n_2-n_3-n_6+n_7+n_8\right)
        \nonumber \\
    & \times
    \Gamma \left(-d+D/2-e-n_3+n_7+n_8+n_9\right)
    \Gamma \left(-D/2+e+f-n_4-n_5-n_6-n_9\right)
        \nonumber \\
    & \times
    \Gamma \left(D/2-f+n_1+n_2+n_3+n_4+n_5+n_6+n_9\right)
\end{align}
\begin{align}
   D_{142} &= (w_6)^{-c-d+D-e-l} (w_7)^{b-D/2+l} (w_8)^{-b}  \sum_{n_1=0}^{\infty} ... \sum_{n_9=0}^{\infty}
    \frac{(-w_1 w_7/w_8)^{n_1}}{n_1!} 
    \frac{(-w_2/w_6)^{n_2}}{n_2!}
    \frac{(-w_3/w_6)^{n_3}}{n_3!}
        \nonumber \\
    & \times
    \frac{(-w_4)^{n_4}}{n_4!}    
    \frac{(-w_5 w_7/w_8)^{n_5}}{n_5!}    
    \frac{(-w_6 w_9/w_8)^{n_6}}{n_6!}
    \frac{(-w_6 w_7/w_8)^{n_7}}{n_7!}
    \frac{(-w_6)^{n_8}}{n_8!}
    \frac{(-w_7)^{n_9}}{n_9!}
    \nonumber \\
    & \times
    \frac{\Gamma \left(b+n_1+n_5+n_6+n_7\right)\Gamma \left(-c-d+3D/2-e-f-l+n_1+n_4+n_5+n_6+n_7+n_8+n_9\right)}{\Gamma \left(D/2-l-n_9\right)}
        \nonumber \\
    & \times
    \Gamma \left(-c+D-e-l-n_2+n_4+n_5+n_6+n_7+n_8\right)
    \Gamma \left(-d+D/2-e-n_3+n_7+n_8+n_9\right)     
        \nonumber \\
    & \times    
    \Gamma \left(c+d-D+e+l+n_2+n_3-n_6-n_7-n_8\right)
    \Gamma \left(-b+D/2-l-n_1-n_5-n_7-n_9\right)    
        \nonumber \\
    & \times    
    \Gamma \left(c+d+2 e+f+l+n_2+n_3-n_4-n_5-n_6-n_7-n_8-n_9-3D/2\right)
\end{align}
\begin{align}
   D_{210} &= (w_5)^{-b-d+D-e-l} (w_8)^{-c} (w_9)^{c-D/2+l} 
   \sum_{n_1=0}^{\infty} ... \sum_{n_9=0}^{\infty}
    \frac{(-w_1/w_5)^{n_1}}{n_1!}
    \frac{(-w_2 w_9/w_8)^{n_2}}{n_2!}
    \frac{(-w_3/w_5)^{n_3}}{n_3!}
        \nonumber \\
    & \times
    \frac{(-w_4)^{n_4}}{n_4!}
    \frac{(-w_6 w_9/w_8)^{n_5}}{n_5!}
    \frac{(-w_5 w_7/w_8)^{n_6}}{n_6!}
    \frac{(-w_5 w_9/w_8)^{n_7}}{n_7!}
    \frac{(-w_5)^{n_8}}{n_8!}
    \frac{(-w_9)^{n_9}}{n_9!}    
        \nonumber \\
    & \times
    \frac{\Gamma \left(c+n_2+n_5+n_6+n_7\right)}{\Gamma \left(D/2-l-n_9\right)}
    \Gamma \left(-b-d+3D/2-e-f-l+n_2+n_4+n_5+n_6+n_7+n_8+n_9\right)
        \nonumber \\
    & \times
    \Gamma \left(b+d-D+e+l+n_1+n_3-n_6-n_7-n_8\right)
        \Gamma \left(-c+D/2-l-n_2-n_5-n_7-n_9\right)
        \nonumber \\
    & \times    
    \Gamma \left(b+d+2 e+f+l+n_1+n_3-n_4-n_5-n_6-n_7-n_8-n_9-3D/2\right)
        \nonumber \\
    & \times    
        \Gamma \left(-b+D-e-l-n_1+n_4+n_5+n_6+n_7+n_8\right)
        \Gamma \left(-d+D/2-e-n_3+n_7+n_8+n_9\right)    
\end{align}
\begin{align}
   D_{826} &= (w_3)^{-d+D/2-e} (w_7)^{b-D/2+l} (w_8)^{-b-c+D/2-l} (w_9)^{c-D/2+l}
    \sum_{n_1=0}^{\infty} ... \sum_{n_9=0}^{\infty}
    \frac{(-w_1 w_7/w_8)^{n_1}}{n_1!}
       \nonumber \\
    & \times
    \frac{(-w_2 w_9/w_8)^{n_2}}{n_2!}
    \frac{(-w_4)^{n_3}}{n_3!}    
    \frac{(-w_5 w_7/w_8)^{n_4}}{n_4!}
    \frac{(-w_6 w_9/w_8)^{n_5}}{n_5!}
    \frac{(-w_3 w_7/w_8)^{n_6}}{n_6!}
        \nonumber \\
    & \times
    \frac{(-w_3 w_9/w_8)^{n_7}}{n_7!}     
    \frac{(-w_3)^{n_8}}{n_8!}
    \frac{(-w_3 w_7 w_9/w_8)^{n_9}}{n_9!}
    \frac{\Gamma \left(-b+D/2-l-n_1-n_4-n_6-n_9\right)}
    {\Gamma \left(D/2-l-n_9\right)}
        \nonumber \\
    & \times
    \Gamma \left(D/2-e+n_3+n_4+n_5+n_6+n_7+n_8+n_9\right)
    \Gamma \left(-D/2+e+f-n_3-n_4-n_5-n_9\right)
        \nonumber \\
    & \times    
    \Gamma \left(-c+D/2-l-n_2-n_5-n_7-n_9\right)
    \Gamma \left(d+e-n_6-n_7-n_8-n_9-D/2\right)
        \nonumber \\
    & \times
    \Gamma \left(-d+D-e-f+n_1+n_2+n_3+n_4+n_5+n_6+n_7+n_8+2 n_9\right)
        \nonumber \\
    & \times
    \Gamma \left(b+c+l+n_1+n_2+n_4+n_5+n_6+n_7+n_9-D/2\right)
\end{align}

\begin{align}
   D_{926} &= (w_3)^{-d+D/2-e} (w_6)^{-c+D/2-l} (w_7)^{b-D/2+l} (w_8)^{-b}
    \sum_{n_1=0}^{\infty} ... \sum_{n_9=0}^{\infty}
    \frac{(-w_1 w_7/w_8)^{n_1}}{n_1!}
    \frac{(-w_2/w_6)^{n_2}}{n_2!}    
       \nonumber \\
    & \times
    \frac{(-w_4)^{n_3}}{n_3!}    
    \frac{(-w_5 w_7/w_8)^{n_4}}{n_4!}
    \frac{(-w_6 w_9/w_8)^{n_5}}{n_5!}
    \frac{(-w_3 w_7/w_8)^{n_6}}{n_6!}
    \frac{(-w_3/w_6)^{n_7}}{n_7!}     
    \frac{(-w_3)^{n_8}}{n_8!}
        \nonumber \\
    & \times
    \frac{(-w_3 w_7/w_6)^{n_9}}{n_9!} 
    \frac{\Gamma \left(b+n_1+n_4+n_5+n_6\right)}
    {\Gamma \left(D/2-l-n_9\right)}
    \Gamma \left(-b+D/2-l-n_1-n_4-n_6-n_9\right)
        \nonumber \\
    & \times
    \Gamma \left(d+e-n_6-n_7-n_8-n_9-D/2\right)
    \Gamma \left(c-D+e+f+l+n_2-n_3-n_4-n_5+n_7\right)    
        \nonumber \\
    & \times    
    \Gamma \left(-c+D-e-l-n_2+n_3+n_4+n_5+n_6+n_8\right)
    \Gamma \left(c+l+n_2-n_5+n_7+n_9-D/2\right)    
        \nonumber \\
    & \times
    \Gamma \left(-c-d+3D/2-e-f-l+n_1+n_3+n_4+n_5+n_6+n_8+n_9\right)
\end{align}

\begin{align}
   D_{942} &= (w_3)^{-d+D/2-e} (w_6)^{-D/2+e+f} (w_7)^{b-D/2+1} (w_8)^{-b-c+D-e-f-l} (w_9)^{c-D+e+f+l}
    \sum_{n_1=0}^{\infty} ... \sum_{n_9=0}^{\infty}
    \nonumber \\
    & \times
    \frac{(-w_1 w_7/w_8)^{n_1}}{n_1!}
    \frac{(-w_2 w_9/w_8)^{n_2}}{n_2!}
    \frac{(-w_4 w_8/(w_6 w_9))^{n_3}}{n_3!}
    \frac{(-w_5 w_7/(w_6 w_9))^{n_4}}{n_4!}
    \frac{(-w_3 w_7/w_8)^{n_5}}{n_5!}
        \nonumber \\
    & \times
    \frac{(-w_3 w_9/w_8)^{n_6}}{n_6!}
    \frac{(-w_6 w_9/w_8)^{n_7}}{n_7!}
    \frac{(-w_3)^{n_8}}{n_8!}
    \frac{(-w_3 w_7/w_6)^{n_9}}{n_9!} 
    \frac{\Gamma \left(f+n_5+n_6+n_7+n_8\right)}
    {\Gamma \left(D/2-l-n_9\right)}
        \nonumber \\
    & \times
    \Gamma \left(-d+D/2+n_1+n_2+n_5+n_6+n_7+n_8+n_9\right)
    \Gamma \left(-b+D/2-l-n_1-n_4-n_5-n_9\right)
        \nonumber \\
    & \times    
    \Gamma \left(d+e-n_5-n_6-n_8-n_9-D/2\right)
    \Gamma \left(D/2-e-f+n_3+n_4-n_7+n_9\right)
        \nonumber \\
    & \times    
    \Gamma \left(-c+D-e-f-l-n_2+n_3+n_4-n_6-n_7\right)
        \nonumber \\
    & \times    
    \Gamma \left(b+c-D+e+f+l+n_1+n_2-n_3+n_5+n_6+n_7\right)
\end{align}
\begin{align}
   D_{988} &= (w_3)^{-d+D/2-e} (w_5)^{-b+D/2-l} (w_8)^{-c} (w_9)^{c-D/2+l}
    \sum_{n_1=0}^{\infty} ... \sum_{n_9=0}^{\infty}
    \frac{(-w_1/w_5)^{n_1}}{n_1!}
    \frac{(-w_2 w_9/w_8)^{n_2}}{n_2!}
    \nonumber \\
        & \times
    \frac{(-w_4)^{n_3}}{n_3!}      
    \frac{(-w_6 w_9/w_8)^{n_4}}{n_4!}
    \frac{(-w_5 w_7/w_8)^{n_5}}{n_5!}
    \frac{(-w_3/w_5)^{n_6}}{n_6!}
    \frac{(-w_3 w_9/w_8)^{n_7}}{n_7!}
    \frac{(-w_3)^{n_8}}{n_8!}
    \frac{(-w_3 w_9/w_5)^{n_9}}{n_9!}
        \nonumber \\
    & \times    
    \frac{\Gamma \left(c+n_2+n_4+n_5+n_7\right)}
    {\Gamma \left(D/2-l-n_9\right))}
    \Gamma \left(-b+D-e-l-n_1+n_3+n_4+n_5+n_7+n_8\right)
        \nonumber \\
    & \times
    \Gamma \left(b+l+n_1-n_5+n_6+n_9-D/2\right)
    \Gamma \left(d+e-n_6-n_7-n_8-n_9-D/2\right)
        \nonumber \\
    & \times     
    \Gamma \left(b-D+e+f+l+n_1-n_3-n_4-n_5+n_6\right)
    \Gamma \left(-c+D/2-l-n_2-n_4-n_7-n_9\right)
        \nonumber \\
    & \times     
    \Gamma \left(-b-d+3D/2-e-f-l+n_2+n_3+n_4+n_5+n_7+n_8+n_9\right)
\end{align}
\begin{align}
   D_{1094} &= (w_3)^{-d+D/2-e} (w_5)^{-b+D/2-l} (w_6)^{-c+D/2-l} (w_8)^{-D/2+l}
    \sum_{n_1=0}^{\infty} ... \sum_{n_9=0}^{\infty}
    \frac{(-w_1/w_5)^{n_1}}{n_1!}
    \nonumber \\
        & \times    
    \frac{(-w_2/w_6)^{n_2}}{n_2!}
    \frac{(-w_4)^{n_3}}{n_3!}
    \frac{(-w_5 w_7/w_8)^{n_4}}{n_4!}    
    \frac{(-w_6 w_9/w_8)^{n_5}}{n_5!}
    \frac{(-w_3/w_5)^{n_6}}{n_6!}
    \frac{(-w_3/w_6)^{n_7}}{n_7!}    
    \nonumber \\
        & \times
    \frac{(-w_3)^{n_8}}{n_8!}        
    \frac{(-w_3/(w_5 w_6 w_8))^{n_9}}{n_9!}
    \frac
    {\Gamma \left(D/2-l+n_4+n_5-n_9\right)}
    {\Gamma \left(D/2-l-n_9\right)}
        \nonumber \\
    & \times
    \Gamma \left(c+l+n_2-n_5+n_7+n_9-D/2\right)
    \Gamma \left(-b-c-d+2D-e-f-2 l+n_3+n_4+n_5+n_8\right)
        \nonumber \\
    & \times
    \Gamma \left(d+e-n_6-n_7-n_8-n_9-D/2\right)
    \Gamma \left(b+l+n_1-n_4+n_6+n_9-D/2\right)     
        \nonumber \\
    & \times    
    \Gamma \left(-b-c+3D/2-e-2 l-n_1-n_2+n_3+n_4+n_5+n_8-n_9\right)
        \nonumber \\
    & \times
    \Gamma \left(b+c+e+f+2 l+n_1+n_2-n_3-n_4-n_5+n_6+n_7+n_9-3 D/2\right)
\end{align}

\begin{align}
   D_{1112} &= (w_3)^{-d+D/2-e} (w_5)^{-b+D/2-l} (w_6)^{b-D+e+f+l} (w_8)^{-b-c+D-e-f-l} (w_9)^{b+c-3D/2+e+f+2l}
    \nonumber \\
        & \times
    \sum_{n_1=0}^{\infty} ... \sum_{n_9=0}^{\infty}    
    \frac{(-w_1 w_6 w_9/(w_5 w_8))^{n_1}}{n_1!}
    \frac{(-w_2 w_9/w_8)^{n_2}}{n_2!}
    \frac{(-w_4 w_8/(w_6 w_9))^{n_3}}{n_3!}
    \frac{(-w_5 w_7/(w_6 w_9))^{n_4}}{n_4!}
    \nonumber \\
        & \times    
    \frac{(-w_3 w_6 w_9/(w_5 w_8))^{n_5}}{n_5!}
    \frac{(-w_3 w_9/w_8)^{n_6}}{n_6!}
    \frac{(-w_6 w_9/w_8)^{n_7}}{n_7!}
    \frac{(-w_3)^{n_8}}{n_8!}
    \frac{(-w_3 w_9/w_5)^{n_9}}{n_9!}
        \nonumber \\
    & \times
    \frac{\Gamma \left(f+n_5+n_6+n_7+n_8\right)}
    {\Gamma \left(D/2-l-n_9\right)}
    \Gamma \left(b+c-D+e+f+l+n_1+n_2-n_3+n_5+n_6+n_7\right)    
        \nonumber \\
    & \times
    \Gamma \left(-d+D/2+n_1+n_2+n_5+n_6+n_7+n_8+n_9\right)
    \Gamma \left(b+l+n_1-n_4+n_5+n_9-D/2\right)
        \nonumber \\
    & \times    
    \Gamma \left(d+e-n_5-n_6-n_8-n_9-D/2\right)
    \Gamma \left(-b+D-e-f-l-n_1+n_3+n_4-n_5-n_7\right)
        \nonumber \\
    & \times    
    \Gamma \left(-b-c+3D/2-e-f-2 l-n_1-n_2+n_3+n_4-n_5-n_6-n_7-n_9\right)
\end{align}
\begin{align}
   D_{1330} &= (w_3)^{-d+D/2-e} (w_4)^{b+c-D+e+f+l} (w_6)^{-b-c+D/2-l} (w_7)^{b-D/2+l} (w_9)^{-b}
   \nonumber \\
        & \times
    \sum_{n_1=0}^{\infty} ... \sum_{n_9=0}^{\infty}    
    \frac{(-w_1 w_4 w_7/(w_6 w_9))^{n_1}}{n_1!}
    \frac{(-w_2 w_4/w_6)^{n_2}}{n_2!}
    \frac{(-w_2 w_5 w_7/(w_6 w_9))^{n_3}}{n_3!}
    \nonumber \\
        & \times
    \frac{(-w_4 w_8/(w_6 w_9))^{n_4}}{n_4!}    
    \frac{(-w_3 w_4 w_7/(w_6 w_9))^{n_5}}{n_5!}
    \frac{(-w_3 w_4/w_6)^{n_6}}{n_6!}
    \frac{(-w_4)^{n_7}}{n_7!}
    \frac{(-w_3)^{n_8}}{n_8!}
    \frac{(-w_3 w_7/w_6)^{n_9}}{n_9!}
        \nonumber \\
    & \times
    \frac{\Gamma \left(b+n_1+n_3+n_4+n_5\right)}
    {\Gamma \left(D/2-l-n_9\right)}
    \Gamma \left(-b-c+D-e-f-l-n_1-n_2-n_4-n_5-n_6-n_7\right)
        \nonumber \\
    & \times
    \Gamma \left(f+n_5+n_6+n_7+n_8\right)
    \Gamma \left(-d+D/2+n_1+n_2+n_5+n_6+n_7+n_8+n_9\right)
        \nonumber \\
    & \times    
    \Gamma \left(-b+D/2-l-n_1-n_3-n_5-n_9\right)
    \Gamma \left(d+e-n_5-n_6-n_8-n_9-D/2\right)    
        \nonumber \\
    & \times
    \Gamma \left(b+c+l+n_1+n_2+n_3+n_4+n_5+n_6+n_9-D/2\right)
\end{align}
\begin{align}
   D_{1449} &= (w_3)^{-d+D/2-e} (w_4)^{b+c-D+e+f+l} (w_5)^{-b+D/2-l} (w_6)^{-c} (w_9)^{-D/2+l}
    \sum_{n_1=0}^{\infty} ... \sum_{n_9=0}^{\infty}    
    \frac{(-w_1 w_4/w_5)^{n_1}}{n_1!}
       \nonumber \\
        & \times
    \frac{(-w_2 w_4/w_6)^{n_2}}{n_2!}    
    \frac{(-w_5 w_7/(w_6 w_9))^{n_3}}{n_3!}
    \frac{(-w_4 w_8/(w_6 w_9))^{n_4}}{n_4!}    
    \frac{(-w_3 w_4/w_5)^{n_5}}{n_5!}
    \frac{(-w_3 w_4/w_6)^{n_6}}{n_6!}
        \nonumber \\
    & \times
    \frac{(-w_4)^{n_7}}{n_7!}    
    \frac{(-w_3)^{n_8}}{n_8!}
    \frac{(-w_3 w_9/w_5)^{n_9}}{n_9!}    
    \frac{\Gamma \left(c+n_2+n_3+n_4+n_6\right)}{\Gamma \left(D/2-l-n_9\right)}
    \Gamma \left(f+n_5+n_6+n_7+n_8\right)    
        \nonumber \\
    & \times
    \Gamma \left(-d+D/2+n_1+n_2+n_5+n_6+n_7+n_8+n_9\right)
    \Gamma \left(D/2-l+n_3+n_4-n_9\right)
        \nonumber \\
    & \times
    \Gamma \left(b+l+n_1-n_3+n_5+n_9-D/2\right)
    \Gamma \left(d+e-n_5-n_6-n_8-n_9-D/2\right)
        \nonumber \\
    & \times
    \Gamma \left(-b-c+D-e-f-l-n_1-n_2-n_4-n_5-n_6-n_7\right)
\end{align}
\begin{align}
   D_{1647} &= (w_2)^{-c+D-e-l} (w_6)^{-d} (w_7)^{b-D/2+l} (w_8)^{-b}
    \sum_{n_1=0}^{\infty} ... \sum_{n_9=0}^{\infty}    
    \frac{(-w_1 w_7/w_8 )^{n_1}}{n_1!}
       \nonumber \\
        & \times
    \frac{(-w_3/w_6)^{n_2}}{n_2!}    
    \frac{(-w_2 w_4/w_6)^{n_3}}{n_3!}
    \frac{(-w_2 w_5 w_7/(w_6 w_8))^{n_4}}{n_4!}    
    \frac{(-w_2 w_9/w_8)^{n_5}}{n_5!}
    \frac{(-w_2 w_7/w_8)^{n_6}}{n_6!}
        \nonumber \\
    & \times
    \frac{(-w_2/w_6)^{n_7}}{n_7!}    
    \frac{(-w_2)^{n_8}}{n_8!}
    \frac{(-w_7)^{n_9}}{n_9!}    
    \frac{\Gamma \left(b+n_1+n_4+n_5+n_6\right)}{\Gamma \left(D/2- l-n_9\right)}
    \Gamma \left(d+n_2+n_3+n_4+n_7\right)
        \nonumber \\
    & \times
   \Gamma \left(-b+D/2-l-n_1-n_4-n_6-n_9\right) \Gamma \left(c-D+e+l-n_3-n_4-n_5-n_6-n_7-n_8\right)
        \nonumber \\
    & \times
    \Gamma \left(-d+D/2-e-n_2+n_6+n_8+n_9\right) \Gamma \left(d+e+f+n_2+n_7-n_9-D/2\right)
        \nonumber \\
    & \times
   \Gamma \left(-c-d+3D/2-e-f-l+n_1+n_3+n_4+n_5+n_6+n_8+n_9\right)
\end{align}
\begin{align}
   D_{2436} &= (w_2)^{-c+D-e-l} (w_3)^{-d+D/2-e} (w_6)^{-D/2+e}
   (w_7)^{b-D/2+l} (w_8)^{-b}
    \sum_{n_1=0}^{\infty} ... \sum_{n_9=0}^{\infty}    
    \frac{(-w_1 w_7/ w_8 )^{n_1}}{n_1!}
       \nonumber \\
        & \times
    \frac{(-w_2 w_4/w_6)^{n_2}}{n_2!}    
    \frac{(-w_2 w_5 w_7/(w_6 w_8))^{n_3}}{n_3!}
    \frac{(-w_2 w_9/w_8)^{n_4}}{n_4!}    
    \frac{(-w_2 w_3 w_7/w_6)^{n_5}}{n_5!}
    \frac{(-w_3 /w_6)^{n_6}}{n_6!}
        \nonumber \\
    & \times
    \frac{(-w_2/w_6)^{n_7}}{n_7!}    
    \frac{(-w_2 w_3/w_6)^{n_8}}{n_8!}
    \frac{(- w_3 w_7 / w_6)^{n_9}}{n_9!}    
    \frac{\Gamma \left(b+n_1+n_3+n_4+n_5\right)}{\Gamma \left(D/2-l-n_9\right)}
        \nonumber \\
    & \times
   \Gamma \left(D/2-e+n_2+n_3+n_5+n_6+n_7+n_8+n_9\right) \Gamma
   \left(-b+D/2-l-n_1-n_3-n_5-n_9\right)
        \nonumber \\
    & \times
    \Gamma \left(d-D/2+e-n_5-n_6-n_8-n_9\right) \Gamma
   \left(c-D+e+l-n_2-n_3-n_4-n_5-n_7-n_8\right) 
        \nonumber \\
    & \times
   \Gamma \left(-c-d+3D/2-e-f-l+n_1+n_2+n_3+n_4+n_5+n_8+n_9\right)\Gamma \left(f+n_5+n_6+n_7+n_8\right)
\end{align}
\begin{align}
   D_{3069} &= (w_1)^{-b+D-e-l} (w_5)^{-d} (w_8)^{-c} (w_9)^{c-D/2+l}
    \sum_{n_1=0}^{\infty} ... \sum_{n_9=0}^{\infty}    
    \frac{(-w_2 w_9/ w_8 )^{n_1}}{n_1!}
       \nonumber \\
        & \times
    \frac{(-w_3 /w_5)^{n_2}}{n_2!}    
    \frac{(-w_1 w_4 /w_5)^{n_3}}{n_3!}
    \frac{(-w_1 w_6 w_9/(w_5 w_8))^{n_4}}{n_4!}    
    \frac{(-w_1 w_7/w_8)^{n_5}}{n_5!}
    \frac{(-w_1 w_9/w_8)^{n_6}}{n_6!}
        \nonumber \\
    & \times
    \frac{(-w_1 /w_5)^{n_7}}{n_7!}    
    \frac{(-w_1)^{n_8}}{n_8!}
    \frac{(-w_9)^{n_9}}{n_9!}    
    \frac{\Gamma \left(c+n_1+n_4+n_5+n_6\right)}{\Gamma \left(D/2-l-n_9\right)}
   \Gamma \left(d+n_2+n_3+n_4+n_7\right)
        \nonumber \\
    & \times
   \Gamma \left(-c+D/2-l-n_1-n_4-n_6-n_9\right) \Gamma
   \left(-d+D/2-e-n_2+n_6+n_8+n_9\right)
        \nonumber \\
    & \times
   \Gamma \left(b-D+e+l-n_3-n_4-n_5-n_6-n_7-n_8\right) \Gamma
   \left(d-D/2+e+f+n_2+n_7-n_9\right)
        \nonumber \\
    & \times
  \Gamma \left(-b-d+3D/2-e-f-l+n_1+n_3+n_4+n_5+n_6+n_8+n_9\right)
\end{align}
\begin{align}
   D_{3806} &= (w_1)^{-b+D-e-l} (w_3)^{-d+D/2-e}(w_5)^{-D/2+e} (w_8)^{-c} (w_9)^{c-D/2+l}
    \sum_{n_1=0}^{\infty} ... \sum_{n_9=0}^{\infty}    
    \frac{(-w_2 w_9 /w_8 )^{n_1}}{n_1!}
       \nonumber \\
        & \times
    \frac{(-w_1 w_4/w_5)^{n_2}}{n_2!}    
    \frac{(-w_1 w_6 w_9/(w_5 w_8))^{n_3}}{n_3!}
    \frac{(-w_1 w_7/w_8)^{n_4}}{n_4!}    
    \frac{(-w_3 /w_5)^{n_5}}{n_5!}
    \frac{(-w_1 w_3 w_9/(w_5 w_8))^{n_6}}{n_6!}
        \nonumber \\
    & \times
    \frac{(-w_1 /w_5)^{n_7}}{n_7!}    
    \frac{(-w_1 w_3/w_5)^{n_8}}{n_8!}
    \frac{(-w_3 w_9 /w_5)^{n_9}}{n_9!}    
    \frac{\Gamma \left(c+n_1+n_3+n_4+n_6\right) }{\Gamma \left(D/2-l-n_9\right)}
        \nonumber \\
    & \times
   \Gamma \left(D/2-e+n_2+n_3+n_5+n_6+n_7+n_8+n_9\right) \Gamma
   \left(-c+D/2-l-n_1-n_3-n_6-n_9\right)
        \nonumber \\
    & \times
   \Gamma \left(d-D/2+e-n_5-n_6-n_8-n_9\right) \Gamma
   \left(b-D+e+l-n_2-n_3-n_4-n_6-n_7-n_8\right)
        \nonumber \\
    & \times
  \Gamma \left(-b-d+3D/2-e-f-l+n_1+n_2+n_3+n_4+n_6+n_8+n_9\right) \Gamma \left(f+n_5+n_6+n_7+n_8\right)
\end{align}

\bigskip

\bigskip

\newpage

{\bf APPENDIX B The hexagon}

\bigskip

\bigskip

We give here the explicit expressions of the 26 series that constitute the series representation of the hexagon in Eq.(\ref{results_I6}).

\begin{align}
   H_1 &= \sum_{n_1=0}^{\infty} ... \sum_{n_9=0}^{\infty}
    \frac{(-w_1)^{n_1}}{n_1!} \frac{(-w_2)^{n_2}}{n_2!} \frac{(-w_3)^{n_3}}{n_3!} \frac{(-w_4)^{n_4}}{n_4!} \frac{(-w_5)^{n_5}}{n_5!} \frac{(-w_6)^{n_6}}{n_6!} \frac{(-w_7)^{n_7}}{n_7!} \frac{(-w_8)^{n_8}}{n_8!} 
    \frac{(-w_9)^{n_9}}{n_9!} \nonumber \\
    & \times \Gamma \left(D/2-f+n_1+n_2+n_3+n_4+n_5+n_6+n_7+n_8+n_9\right) \nonumber \\
    & \times \Gamma \left(b+n_1+n_5+n_8+n_9\right) 
    \Gamma \left(c+n_2+n_6+n_7+n_8\right)  
    \Gamma \left(d+n_3+n_4+n_5+n_6\right) \nonumber \\
    & \times
    \Gamma \left(D/2-b-c-d-e-n_1-n_2-n_3-n_5-n_6-n_8\right) \nonumber \\
    & \times \Gamma \left(e+f-D/2-n_4-n_5-n_6-n_7-n_8-n_9\right)
\end{align} 

\begin{align}
   H_4 &= (w_9)^{e+f-D/2} \sum_{n_1=0}^{\infty} ... \sum_{n_9=0}^{\infty}
    \frac{(-w_1)^{n_1}}{n_1!} \frac{(-w_2)^{n_2}}{n_2!} \frac{(-w_3)^{n_3}}{n_3!}
    \frac{(-w_4/w_9)^{n_4}}{n_4!} \frac{(-w_5/w_9)^{n_5}}{n_5!} 
    \frac{(-w_6/w_9)^{n_6}}{n_6!} \nonumber \\
    & \times \frac{(-w_7/w_9)^{n_7}}{n_7!} \frac{(-w_8/w_9)^{n_8}}{n_8!} \frac{(-w_9)^{n_9}}{n_9!}
    \Gamma \left(c+n_2+n_6+n_7+n_8\right)
    \Gamma \left(d+n_3+n_4+n_5+n_6\right) \nonumber \\
    & \times
    \Gamma \left(b+e+f-D/2+n_1-n_4-n_6-n_7+n_9\right)
      \Gamma \left(D/2-e-f+n_4+n_5+n_6+n_7+n_8-n_9\right)
    \nonumber \\
    & \times 
    \Gamma \left(-b-c-d+D/2-e-n_1-n_2-n_3-n_5-n_6-n_8\right)
     \Gamma \left( e+n_1+n_2+n_3+n_9\right)
\end{align}

\begin{align}
   H_9 &= (w_8)^{-b-c-d-e+D/2} \sum_{n_1=0}^{\infty} ... \sum_{n_9=0}^{\infty}
    \frac{(-w_1/w_8)^{n_1}}{n_1!} \frac{(-w_2/w_8)^{n_2}}{n_2!} \frac{(-w_3/w_8)^{n_3}}{n_3!} \frac{(-w_4)^{n_4}}{n_4!} \frac{(-w_5/w_8)^{n_5}}{n_5!} \nonumber \\
    & \times \frac{(-w_6/w_8)^{n_6}}{n_6!}\frac{(-w_7)^{n_7}}{n_7!} \frac{(-w_9)^{n_8}}{n_8!} \frac{(-w_8)^{n_9}}{n_9!} 
    \Gamma \left(b+c+d+e+n_1+n_2+n_3+n_5+n_6-n_9-D/2\right)
    \nonumber \\
    & \times
     \Gamma \left(d+n_3+n_4+n_5+n_6\right)\Gamma \left(-b-d+D/2-e-n_1-n_3-n_5+n_7+n_9\right) 
     \nonumber \\ &
    \times \Gamma \left(-c-d+D/2-e-n_2-n_3-n_6+n_8+n_9\right) \Gamma \left(-b-c-d+D-e-f+n_4+n_7+n_8+n_9\right) \nonumber \\ &
    \times \Gamma \left(b+c+d-D+2 e+f+n_1+n_2+n_3-n_4-n_7-n_8-n_9\right)
\end{align}

\begin{align}
   H_{15} &= (w_8)^{-b-c-d-e+D/2} (w_9)^{b+c+d+2e+f-D} \sum_{n_1=0}^{\infty} ... \sum_{n_9=0}^{\infty}
    \frac{(-w_1 w_9/w_8)^{n_1}}{n_1!} \frac{(-w_2 w_9/w_8)^{n_2}}{n_2!} \frac{(-w_3 w_9/w_8)^{n_3}}{n_3!}   \nonumber \\
    & \times \frac{(-w_4/w_9)^{n_4}}{n_4!}\frac{(-w_5/w_8)^{n_5}}{n_5!} \frac{(-w_6/w_8)^{n_6}}{n_6!} \frac{(-w_7/w_9)^{n_7}}{n_7!} \frac{(-w_9)^{n_8}}{n_8!} \frac{(-w_8/w_9)^{n_9}}{n_9!} \nonumber \\
    & \times 
    \Gamma \left(d+n_3+n_4+n_5+n_6\right) \Gamma
   \left(-b-c-d+D-2 e-f-n_1-n_2-n_3+n_4+n_7-n_8+n_9\right) \nonumber \\
    & \times
    \Gamma \left(-b-d+D/2-e-n_1-n_3-n_5+n_7+n_9\right) \Gamma
   \left(b-D/2+e+f+n_1-n_4-n_6-n_7+n_8\right)
    \nonumber \\
    & \times \Gamma \left(b+c+d-D/2+e+n_1+n_2+n_3+n_5+n_6-n_9\right) \Gamma \left(e+n_1+n_2+n_3+n_8\right)
\end{align}

\begin{align}
   H_{22} &=  w_7^{b+e+f-D/2} w_9^{-b} \sum_{n_1=0}^{\infty} ... \sum_{n_9=0}^{\infty}
    \frac{(-w_1 w_7/w_9)^{n_1}}{n_1!} \frac{(-w_2)^{n_2}}{n_2!} \frac{(-w_3)^{n_3}}{n_3!} \frac{(-w_4/w_7)^{n_4}}{n_4!} 
    \frac{(-w_5/w_9)^{n_5}}{n_5!} \nonumber \\
    & \times \frac{(-w_6/w_7)^{n_6}}{n_6!} \frac{(-w_8/w_9)^{n_7}}{n_7!} \frac{(-w_7/w_9)^{n_8}}{n_8!} \frac{(-w_7)^{n_9}}{n_9!}
    \Gamma \left(b+n_1+n_5+n_7+n_8\right)
    \nonumber \\
    & \times  \Gamma \left(-b-c-d+D/2-e-n_1-n_2-n_3-n_5-n_6-n_7\right)  \nonumber \\
    & \times 
    \Gamma
   \left(-b+D/2-e-f-n_1+n_4+n_6-n_8-n_9\right)\Gamma \left(e+n_1+n_2+n_3+n_9\right) \nonumber \\
    & \times 
   \Gamma \left(d+n_3+n_4+n_5+n_6\right)\Gamma
   \left(b+c-D/2+e+f+n_1+n_2-n_4+n_7+n_8+n_9\right)
\end{align}
\begin{align}
   H_{35} &= w_7^{b+e+f-D/2} w_8^{-b-c-d-e+D/2} w_9^{c+d+e-D/2} \sum_{n_1=0}^{\infty} ... \sum_{n_9=0}^{\infty}
    \frac{(-w_1 w_7/w_8)^{n_1}}{n_1!} \frac{(-w_2 w_9/w_8)^{n_2}}{n_2!}  \nonumber \\
    & \times \frac{(-w_3 w_9/w_8)^{n_3}}{n_3!}\frac{(-w_4/w_7)^{n_4}}{n_4!} \frac{(-w_5/w_8)^{n_5}}{n_5!}  \frac{(-w_6 w_9/(w_7 w_8))^{n_6}}{n_6!} \frac{(-w_7/w_9)^{n_7}}{n_7!} \frac{(-w_7)^{n_8}}{n_8!} \frac{(-w_8/w_9)^{n_9}}{n_9!}
    \nonumber \\
    & \times \Gamma \left(d+n_3+n_4+n_5+n_6\right) \Gamma \left(e+n_1+n_2+n_3+n_8\right)
        \nonumber \\
    & \times \Gamma \left(-d+f-n_3-n_4-n_5-n_6+n_7+n_8+n_9\right) \Gamma
   \left(-b+D/2-e-f-n_1+n_4+n_6-n_7-n_8\right)
        \nonumber \\
    & \times \Gamma \left(-c-d+D/2-e-n_2-n_3-n_6+n_7+n_9\right)
        \nonumber \\
    & \times 
   \Gamma
   \left(b+c+d-D/2+e+n_1+n_2+n_3+n_5+n_6-n_9\right)
\end{align}

\begin{align}
   H_{58} &= w_6^{-c-d-e+D/2} w_8^{-b} 
   \sum_{n_1=0}^{\infty} ... \sum_{n_9=0}^{\infty}
    \frac{(-w_1/w_8)^{n_1}}{n_1!} \frac{(-w_2/w_6)^{n_2}}{n_2!}
    \frac{(-w_3/w_6)^{n_3}}{n_3!}
    \frac{(-w_4)^{n_4}}{n_4!} \frac{(-w_5/w_8)^{n_5}}{n_5!}
    \nonumber \\
    & \times 
    \frac{(-w_7)^{n_6}}{n_6!} \frac{(-w_6 w_9/w_8)^{n_7}}{n_7!} \frac{(-w_6/w_8)^{n_8}}{n_8!} \frac{(-w_6)^{n_9}}{n_9!}
    \Gamma \left(b+n_1+n_5+n_7+n_8\right)
    \nonumber \\
    & \times \Gamma \left(-c+D/2-e-n_2+n_4+n_5+n_7+n_8+n_9\right) \Gamma
   \left(-b-d+D/2-e-n_1-n_3-n_5+n_6+n_9\right)
        \nonumber \\
    & \times
    \Gamma \left(c+d-D/2+e+n_2+n_3-n_7-n_8-n_9\right) \Gamma \left(-b-c-d+D-e-f+n_4+n_6+n_7+n_9\right)
        \nonumber \\
    & \times 
   \Gamma
   \left(b+c+d-D+2 e+f+n_1+n_2+n_3-n_4-n_6-n_7-n_9\right)
\end{align}

\begin{align}
   H_{94} &= w_6^{-c-d-e+D/2} w_7^{b+c+d+2e+f-D}  w_8^{-b} 
   \sum_{n_1=0}^{\infty} ... \sum_{n_9=0}^{\infty}
    \frac{(-w_1 w_7/w_8)^{n_1}}{n_1!} \frac{(-w_2 w_7/w_6)^{n_2}}{n_2!}
    \frac{(-w_3 w_7/w_6)^{n_3}}{n_3!}
    \nonumber \\
    & \times  \frac{(-w_4/w_7)^{n_4}}{n_4!}  \frac{(-w_5/w_8)^{n_5}}{n_5!} \frac{(-w_6 w_9/(w_7 w_8))^{n_6}}{n_6!}
    \frac{(-w_6/w_8)^{n_7}}{n_7!} \frac{(-w_7)^{n_8}}{n_8!} 
    \frac{(-w_6/w_7)^{n_9}}{n_9!} 
    \nonumber \\
    & \times
    \Gamma \left(b+n_1+n_5+n_6+n_7\right)\Gamma
   \left(-b-c-d+D-2 e-f-n_1-n_2-n_3+n_4+n_6-n_8+n_9\right)
        \nonumber \\
    & \times
    \Gamma \left(-c+D/2-e-n_2+n_4+n_5+n_6+n_7+n_9\right) \Gamma
   \left(c+d-D/2+e+n_2+n_3-n_6-n_7-n_9\right)
        \nonumber \\
    & \times
    \Gamma \left(c-D/2+e+f+n_2-n_4-n_5-n_6+n_8\right) \Gamma \left(e+n_1+n_2+n_3+n_8\right)
\end{align}

\begin{align}
   H_{103} &= w_6^{-d+f} w_7^{b+d+e-D/2}  w_8^{-b-c+D/2-e-f} w_9^{c+e+f-D/2} 
   \sum_{n_1=0}^{\infty} ... \sum_{n_9=0}^{\infty} 
    \frac{(-w_1 w_7/w_8)^{n_1}}{n_1!}
        \nonumber \\
    & \times
    \frac{(-w_2 w_9/w_8)^{n_2}}{n_2!} \frac{(-w_3 w_7/w_6)^{n_3}}{n_3!}
    \frac{(-w_4 w_8/(w_6 w_9))^{n_4}}{n_4!} \frac{(-w_5 w_7/(w_6 w_9))^{n_5}}{n_5!}
    \frac{(-w_6/w_8)^{n_6}}{n_6!}
        \nonumber \\
    & \times  
    \frac{(-w_6 w_9/(w_7 w_8))^{n_7}}{n_7!} \frac{(-w_6 w_9/w_8)^{n_8}}{n_8!} 
    \frac{(-w_6/w_7)^{n_9}}{n_9!} 
    \Gamma \left( e+n_1+n_2+n_3+n_8\right)
        \nonumber \\
    & \times
   \Gamma \left(b+c-D/2+e+f+n_1+n_2-n_4+n_6+n_7+n_8\right)\Gamma \left(f+n_6+n_7+n_8+n_9\right)
        \nonumber \\
    & \times
    \Gamma \left(d-f+n_3+n_4+n_5-n_6-n_7-n_8-n_9\right)\Gamma \left(-c+D/2-e-f-n_2+n_4+n_5-n_7-n_8\right)
        \nonumber \\
    & \times
    \Gamma \left(-b-d+D/2-e-n_1-n_3-n_5+n_7+n_9\right)
\end{align}

\begin{align}
   H_{123} &= w_5^{-b-d-e+D/2} w_8^{-c} 
   \sum_{n_1=0}^{\infty} ... \sum_{n_9=0}^{\infty}
    \frac{(-w_1/w_5)^{n_1}}{n_1!} \frac{(-w_2/w_8)^{n_2}}{n_2!}
    \frac{(-w_3/w_5)^{n_3}}{n_3!} \frac{(-w_4)^{n_4}}{n_4!}
    \frac{(-w_6/w_8)^{n_5}}{n_5!}
        \nonumber \\
    & \times \frac{(-w_5 w_7/w_8)^{n_6}}{n_6!}
    \frac{(-w_9)^{n_7}}{n_7!} \frac{(-w_5/w_8)^{n_8}}{n_8!} 
    \frac{(-w_5)^{n_9}}{n_9!} 
    \Gamma \left(c+n_2+n_5+n_6+n_8\right)
        \nonumber \\
    & \times
    \Gamma \left(-b+D/2-e-n_1+n_4+n_5+n_6+n_8+n_9\right) \Gamma
   \left(b+d-D/2+e+n_1+n_3-n_6-n_8-n_9\right)
        \nonumber \\
    & \times
    \Gamma \left(-c-d+D/2-e-n_2-n_3-n_5+n_7+n_9\right)
    \Gamma \left(-b-c-d+D-e-f+n_4+n_6+n_7+n_9\right)
        \nonumber \\
    & \times
  \Gamma
   \left(b+c+d-D+2 e+f+n_1+n_2+n_3-n_4-n_6-n_7-n_9\right)
\end{align}

\begin{align}
   H_{133} &= w_5^{-b-d-e+D/2} w_8^{-c} w_9^{b+c+d+2e+f-D} 
   \sum_{n_1=0}^{\infty} ... \sum_{n_9=0}^{\infty}
    \frac{(-w_1 w_9/w_5)^{n_1}}{n_1!} \frac{(-w_2 w_9/w_8)^{n_2}}{n_2!}
    \frac{(-w_3 w_9/w_5)^{n_3}}{n_3!} 
        \nonumber \\
    & \times
    \frac{(-w_4/w_9)^{n_4}}{n_4!}\frac{(-w_6/w_8)^{n_5}}{n_5!}
    \frac{(-w_5 w_7/(w_8 w_9))^{n_6}}{n_6!}
    \frac{(-w_5/w_8)^{n_7}}{n_7!} \frac{(-w_9)^{n_8}}{n_8!} 
    \frac{(-w_5/w_9)^{n_9}}{n_9!}
        \nonumber \\
    & \times
    \Gamma \left(e+n_1+n_2+n_3+n_8\right) \Gamma \left(-b+D/2-e-n_1+n_4+n_5+n_6+n_7+n_9\right)
        \nonumber \\
    & \times
    \Gamma
   \left(b+d-D/2+e+n_1+n_3-n_6-n_7-n_9\right) \Gamma \left(b-D/2+e+f+n_1-n_4-n_5-n_6+n_8\right)
        \nonumber \\
    & \times
    \Gamma
   \left(-b-c-d+D-2 e-f-n_1-n_2-n_3+n_4+n_6-n_8+n_9\right)\Gamma \left(c+n_2+n_5+n_6+n_7\right) 
\end{align}

\begin{align}
   H_{199} &= w_5^{-b-d-e+D/2} w_6^{-c-d-e+D/2} w_8^{d+e-D/2} 
   \sum_{n_1=0}^{\infty} ... \sum_{n_9=0}^{\infty}
    \frac{(-w_1/w_5)^{n_1}}{n_1!} \frac{(-w_2/w_6)^{n_2}}{n_2!}
    \frac{(-w_3 w_8/(w_5 w_6))^{n_3}}{n_3!}
        \nonumber \\
    & \times
    \frac{(-w_4)^{n_4}}{n_4!}
    \frac{(-w_5 w_7/w_8)^{n_5}}{n_5!} \frac{(-w_6 w_9/w_8)^{n_6}}{n_6!}
    \frac{(-w_6/w_8)^{n_7}}{n_7!} \frac{(-w_5/w_8)^{n_8}}{n_8!} 
    \frac{(-w_5 w_6/w_8)^{n_9}}{n_9!}
        \nonumber \\
    & \times
    \Gamma \left(-d+D/2-e-n_3+n_5+n_6+n_7+n_8+n_9\right) \Gamma \left(b+d-D/2+e+n_1+n_3-n_5-n_8-n_9\right)
        \nonumber \\
    & \times
    \Gamma
   \left(c+d-D/2+e+n_2+n_3-n_6-n_7-n_9\right) \Gamma \left(-b-c-d+D-e-f+n_4+n_5+n_6+n_9\right)
        \nonumber \\
    & \times\Gamma
   \left(b+c+d-D+2 e+f+n_1+n_2+n_3-n_4-n_5-n_6-n_9\right)
        \nonumber \\
    & \times
    \Gamma \left(-b-c-d+D-2 e-n_1-n_2-n_3+n_4+n_5+n_6+n_7+n_8+2 n_9\right)
\end{align}

\begin{align}
   H_{210} &= w_5^{-b-d-e+D/2} w_6^{b+e+f-D/2} w_8^{-b-c+D/2-e-f} w_9^{b+c+d+2e+f-D} 
   \sum_{n_1=0}^{\infty} ... \sum_{n_9=0}^{\infty}
    \frac{(-w_1 w_6 w_9/(w_5 w_8))^{n_1}}{n_1!} 
        \nonumber \\
    & \times
    \frac{(-w_2 w_9/w_8)^{n_2}}{n_2!}
    \frac{(-w_3 w_9/w_5)^{n_3}}{n_3!}
    \frac{(-w_4 w_8/(w_6 w_9))^{n_4}}{n_4!} \frac{(-w_5 w_7(w_6 w_9))^{n_5}}{n_5!} \frac{(-w_6/w_8)^{n_6}}{n_6!} 
        \nonumber \\
    & \times
    \frac{(-w_5/(w_6 w_8))^{n_7}}{n_7!}\frac{(-w_6 w_9/w_8)^{n_8}}{n_8!} \frac{(-w_5/w_9)^{n_9}}{n_9!}
   \Gamma \left(e+n_1+n_2+n_3+n_8\right)
        \nonumber \\
    & \times
   \Gamma
   \left(-b+D/2-e-f-n_1+n_4+n_5-n_6-n_8\right)  \Gamma \left(f+n_6+n_7+n_8+n_9\right) \nonumber \\
    & \times \Gamma \left(b+d-D/2+e+n_1+n_3-n_5-n_7-n_9\right)
        \nonumber \\
    & \times
     \Gamma \left(b+c-D/2+e+f+n_1+n_2-n_4+n_6+n_7+n_8\right)
        \nonumber \\
    & \times
   \Gamma
   \left(-b-c-d+D-2 e-f-n_1-n_2-n_3+n_4+n_5-n_8+n_9\right)
\end{align}

\begin{align}
   H_{270} &= w_4^{b+c+e+f-D/2} w_7^{-c} w_9^{-b}
   \sum_{n_1=0}^{\infty} ... \sum_{n_9=0}^{\infty}
    \frac{(-w_1 w_4/w_9)^{n_1}}{n_1!} \frac{(-w_2 w_4/w_7)^{n_2}}{n_2!}
    \frac{(-w_3)^{n_3}}{n_3!} \frac{(-w_5/w_9)^{n_4}}{n_4!}
        \nonumber \\
    & \times
    \frac{(-w_6/w_7)^{n_5}}{n_5!} \frac{(-w_4 w_8/(w_7 w_9))^{n_6}}{n_6!} \frac{(-w_4/w_9)^{n_7}}{n_7!} \frac{(-w_4/w_7)^{n_8}}{n_8!} \frac{(-w_4)^{n_9}}{n_9!}
    \Gamma \left(b+n_1+n_4+n_6+n_7\right)
        \nonumber \\
    & \times
    \Gamma \left(e+n_1+n_2+n_3+n_9\right)\Gamma
   \left(-b-c-d+D/2-e-n_1-n_2-n_3-n_4-n_5-n_6\right)
        \nonumber \\
    & \times
     \Gamma \left(-b-c+D/2-e-f-n_1-n_2-n_6-n_7-n_8-n_9\right)\Gamma \left(c+n_2+n_5+n_6+n_8\right) 
        \nonumber \\
    & \times \Gamma
   \left(b+c+d-D/2+e+f+n_1+n_2+n_3+n_4+n_5+n_6+n_7+n_8+n_9\right)
\end{align}

\begin{align}
   H_{333} &= w_4^{b+c+e+f-D/2} w_6^{-b-c-d-e+D/2} w_7^{b+d+e-D/2} w_9^{-b}
   \sum_{n_1=0}^{\infty} ... \sum_{n_9=0}^{\infty}
    \frac{(-w_1 w_4 w_7/(w_6 w_9))^{n_1}}{n_1!} 
        \nonumber \\
    & \times
    \frac{(-w_2 w_4/w_6)^{n_2}}{n_2!}\frac{(-w_3 w_7/w_6)^{n_3}}{n_3!} \frac{(-w_5 w_7/(w_6 w_9))^{n_4}}{n_4!}
    \frac{(-w_4 w_8/(w_6 w_9))^{n_5}}{n_5!} \frac{(-w_4/w_9)^{n_6}}{n_6!} 
        \nonumber \\
    & \times
    \frac{(-w_4/w_7)^{n_7}}{n_7!} \frac{(-w_4)^{n_8}}{n_8!}\frac{(-w_6/w_7)^{n_9}}{n_9!}
    \Gamma \left(b+n_1+n_4+n_5+n_6\right) \Gamma \left(e+n_1+n_2+n_3+n_8\right)
        \nonumber \\
    & \times
   \Gamma \left(f+n_6+n_7+n_8+n_9\right) \Gamma
   \left(-b-d+D/2-e-n_1-n_3-n_4+n_7+n_9\right)
        \nonumber \\
    & \times
    \Gamma \left(b+c+d-D/2+e+n_1+n_2+n_3+n_4+n_5-n_9\right)   
        \nonumber \\
    & \times
    \Gamma
   \left(-b-c+D/2-e-f-n_1-n_2-n_5-n_6-n_7-n_8\right)
\end{align}

\begin{align}
   H_{409} &= w_4^{b+c+e+f-D/2} w_5^{-b-d-e+D/2} w_6^{-c} w_9^{d+e-D/2}
   \sum_{n_1=0}^{\infty} ... \sum_{n_9=0}^{\infty}
    \frac{(-w_1 w_4/w_5)^{n_1}}{n_1!} \frac{(-w_2 w_4/w_6)^{n_2}}{n_2!}
        \nonumber \\
    & \times
    \frac{(-w_3 w_9/w_5)^{n_3}}{n_3!} \frac{(-w_5 w_7/(w_6 w_9))^{n_4}}{n_4!}
    \frac{(-w_4 w_8/(w_6 w_9))^{n_5}}{n_5!} \frac{(-w_4/w_9)^{n_6}}{n_6!}
    \frac{(-w_4 w_5/(w_6 w_9))^{n_7}}{n_7!}
        \nonumber \\
    & \times 
        \frac{(-w_4)^{n_8}}{n_8!} \frac{(-w_5/w_9)^{n_9}}{n_9!}
        \Gamma \left(c+n_2+n_4+n_5+n_7\right)
    \Gamma \left(e+n_1+n_2+n_3+n_8\right) 
        \nonumber \\
    & \times
    \Gamma
   \left(-d+D/2-e-n_3+n_4+n_5+n_6+n_7+n_9\right) \Gamma \left(b+d-D/2+e+n_1+n_3-n_4-n_7-n_9\right)
        \nonumber \\
    & \times
    \Gamma
   \left(-b-c+D/2-e-f-n_1-n_2-n_5-n_6-n_7-n_8\right)\Gamma \left(f+n_6+n_7+n_8+n_9\right)
\end{align}

\begin{align}
   H_{637} &= w_3^{-d-e+D/2} w_5^{-b} w_6^{-c}
   \sum_{n_1=0}^{\infty} ... \sum_{n_9=0}^{\infty}
    \frac{(-w_1/w_5)^{n_1}}{n_1!} \frac{(-w_2/w_6)^{n_2}}{n_2!}
    \frac{(-w_4)^{n_3}}{n_3!} \frac{(-w_3 w_7/w_6)^{n_4}}{n_4!}
        \nonumber \\
    & \times
    \frac{(-w_3 w_8/(w_5 w_6))^{n_5}}{n_5!} \frac{(-w_3 w_9/w_5)^{n_6}}{n_6!}
    \frac{(-w_3/w_5)^{n_7}}{n_7!} \frac{(-w_3/w_6)^{n_8}}{n_8!} \frac{(-w_3)^{n_9}}{n_9!}
        \nonumber \\
    & \times 
        \Gamma \left(b+n_1+n_5+n_6+n_7\right) \Gamma \left(c+n_2+n_4+n_5+n_8\right)
        \nonumber \\
    & \times 
        \Gamma \left(d-D/2+e-n_4-n_5-n_6-n_7-n_8-n_9\right) 
        \nonumber \\
    & \times 
        \Gamma \left(b+c-D/2+e+f+n_1+n_2-n_3+n_5+n_7+n_8\right) \nonumber \\
    & \times \Gamma
   \left(-b-c-d+D-e-f+n_3+n_4+n_6+n_9\right) \nonumber \\
    & \times \Gamma
   \left(-b-c+D/2-e-n_1-n_2+n_3-n_5+n_9\right)
\end{align}

\begin{align}
   H_{653} &= w_3^{D-b-c-d-2e} w_5^{c+e-D/2} w_6^{b+e-D/2} w_8^{-b-c-e+D/2}
   \sum_{n_1=0}^{\infty} ... \sum_{n_9=0}^{\infty}
    \frac{(-w_1 w_6/(w_3 w_8))^{n_1}}{n_1!} \nonumber \\
    & \times 
    \frac{(-w_2 w_5/(w_3 w_8))^{n_2}}{n_2!}
    \frac{(-w_3 w_4 w_8/(w_5w_6))^{n_3}}{n_3!} \frac{(-w_3 w_7/w_6)^{n_4}}{n_4!}
    \frac{(-w_3 w_9/w_5)^{n_5}}{n_5!} \frac{(-w_3/w_5)^{n_6}}{n_6!} 
        \nonumber \\
    & \times
   \frac{(-w_3/w_6)^{n_7}}{n_7!} \frac{(-w_3 w_8/(w_5 w_6))^{n_8}}{n_8!} \frac{(-w_3^2 w_8/(w_5 w_6))^{n_9}}{n_9!}
    \Gamma \left(f+n_6+n_7+n_8+n_9\right)
        \nonumber \\
    & \times
        \Gamma \left(-b+D/2-e-n_1+n_3+n_4+n_7+n_8+n_9\right) \nonumber \\
    & \times \Gamma
   \left(-c+D/2-e-n_2+n_3+n_5+n_6+n_8+n_9\right)
        \nonumber \\
    & \times
       \Gamma \left(b+c-D/2+e+n_1+n_2-n_3-n_8-n_9\right) \nonumber \\
    & \times\Gamma \left(-b-c-d+D-e-f+n_3+n_4+n_5+n_9\right)
       \nonumber \\
    & \times    
        \Gamma
   \left(b+c+d-D+2 e+n_1+n_2-n_3-n_4-n_5-n_6-n_7-n_8-2 n_9\right)
\end{align}

\begin{align}
   H_{838} &= w_3^{-d-e+D/2} w_4^{b+c+e+f-D/2} w_5^{-b} w_6^{-c}
   \sum_{n_1=0}^{\infty} ... \sum_{n_9=0}^{\infty}
    \frac{(-w_1 w_4/w_5)^{n_1}}{n_1!}
    \frac{(-w_2 w_4/w_6)^{n_2}}{n_2!}
    \frac{(-w_3 w_7/w_6)^{n_3}}{n_3!} 
    \nonumber \\
    & \times
    \frac{(-w_3 w_4 w_8/(w_5 w_6))^{n_4}}{n_4!}
    \frac{(-w_3 w_9/w_5)^{n_5}}{n_5!}
    \frac{(-w_3 w_4/w_5)^{n_6}}{n_6!}
    \frac{(-w_3 w_4/w_6)^{n_7}}{n_7!}
    \frac{(-w_4)^{n_8}}{n_8!}
    \frac{(-w_3)^{n_9}}{n_9!}
        \nonumber \\
    & \times
       \Gamma \left(b+n_1+n_4+n_5+n_6\right) \Gamma \left(c+n_2+n_3+n_4+n_7\right)
        \nonumber \\
    & \times
       \Gamma \left(f+n_6+n_7+n_8+n_9\right) \Gamma
   \left(-d+D/2+n_1+n_2+n_3+n_4+n_5+n_6+n_7+n_8+n_9\right)
        \nonumber \\
    & \times
        \Gamma \left(d-D/2+e-n_3-n_4-n_5-n_6-n_7-n_9\right)
        \nonumber \\
    & \times
       \Gamma
   \left(-b-c+D/2-e-f-n_1-n_2-n_4-n_6-n_7-n_8\right)
\end{align}

\begin{align}
   H_{925} &= w_2^{-c-e+D/2} w_6^{-d} w_8^{-b}
   \sum_{n_1=0}^{\infty} ... \sum_{n_9=0}^{\infty}
    \frac{(-w_1/w_8)^{n_1}}{n_1!}
    \frac{(-w_3/w_6)^{n_2}}{n_2!}
    \frac{(-w_2 w_4/w_6)^{n_3}}{n_3!}
    \frac{(-w_2 w_5/(w_6 w_8))^{n_4}}{n_4!}    
    \nonumber \\
    & \times
    \frac{(-w_7)^{n_5}}{n_5!}
    \frac{(-w_2 w_9/w_8)^{n_6}}{n_6!}
    \frac{(-w_2/w_8)^{n_7}}{n_7!}
    \frac{(-w_2/w_6)^{n_8}}{n_8!}
    \frac{(-w_2)^{n_9}}{n_9!}
    \Gamma \left(b+n_1+n_4+n_6+n_7\right)
        \nonumber \\
    & \times
        \Gamma \left(d+n_2+n_3+n_4+n_8\right) \Gamma \left(c-D/2+e-n_3-n_4-n_6-n_7-n_8-n_9\right)      
        \nonumber \\
    & \times    
        \Gamma
   \left(-b-d+D/2-e-n_1-n_2-n_4+n_5+n_9\right)\Gamma
   \left(-b-c-d+D-e-f+n_3+n_5+n_6+n_9\right) 
        \nonumber \\
    & \times    
        \Gamma \left(b+d-D/2+e+f+n_1+n_2+n_4-n_5+n_7+n_8\right)
\end{align}

\begin{align}
   H_{960} &= w_2^{-c-e+D/2} w_6^{-d} w_7^{b+d+e+f-D/2} w_8^{-b}
   \sum_{n_1=0}^{\infty} ... \sum_{n_9=0}^{\infty}
    \frac{(-w_1 w_7/w_8)^{n_1}}{n_1!}
    \frac{(-w_3 w_7/w_6)^{n_2}}{n_2!}
    \frac{(-w_2 w_4/w_6)^{n_3}}{n_3!}
    \nonumber \\
    & \times
    \frac{(-w_2 w_5 w_7/(w_6 w_8))^{n_4}}{n_4!}    
    \frac{(-w_2 w_9/w_8)^{n_5}}{n_5!}
    \frac{(-w_2 w_7/w_8)^{n_6}}{n_6!}
    \frac{(-w_2 w_7/w_6)^{n_7}}{n_7!}
    \frac{(-w_7)^{n_8}}{n_8!}
    \frac{(-w_2)^{n_9}}{n_9!}
        \nonumber \\
    & \times
        \Gamma \left(b+n_1+n_4+n_5+n_6\right) \Gamma \left(d+n_2+n_3+n_4+n_7\right)
        \nonumber \\
    & \times 
        \Gamma \left(f+n_6+n_7+n_8+n_9\right) \Gamma
   \left(-c+D/2+n_1+n_2+n_3+n_4+n_5+n_6+n_7+n_8+n_9\right)
        \nonumber \\
    & \times     
        \Gamma \left(c-D/2+e-n_3-n_4-n_5-n_6-n_7-n_9\right)
        \nonumber \\
    & \times      
        \Gamma
   \left(-b-d+D/2-e-f-n_1-n_2-n_4-n_6-n_7-n_8\right)
\end{align}

\begin{align}
   H_{1375} &= w_2^{-c-e+D/2} w_3^{-b-d-e+D/2} w_6^{e+b-D/2} w_8^{-b}
   \sum_{n_1=0}^{\infty} ... \sum_{n_9=0}^{\infty}
    \frac{(-w_1 w_6/(w_3 w_8))^{n_1}}{n_1!}
    \frac{(-w_2 w_4/w_6)^{n_2}}{n_2!}
    \nonumber \\
    & \times
    \frac{(-w_2 w_5/(w_3 w_8))^{n_3}}{n_3!}
    \frac{(-w_3 w_7/w_6)^{n_4}}{n_4!}    
    \frac{(-w_2 w_9/w_8)^{n_5}}{n_5!}
    \frac{(-w_2/w_8)^{n_6}}{n_6!}
    \frac{(-w_3/w_6)^{n_7}}{n_7!}
    \frac{(-w_2/w_6)^{n_8}}{n_8!}
        \nonumber \\
    & \times
        \frac{(-w_2 w_3/w_6)^{n_9}}{n_9!}
        \Gamma \left(b+n_1+n_3+n_5+n_6\right)  \Gamma \left(-b+D/2-e-n_1+n_2+n_4+n_7+n_8+n_9\right)  
        \nonumber \\
    & \times   
       \Gamma \left(f+n_6+n_7+n_8+n_9\right) \Gamma
   \left(-b-c-d+D-e-f+n_2+n_4+n_5+n_9\right)
        \nonumber \\
    & \times
      \Gamma \left(b+d-D/2+e+n_1+n_3-n_4-n_7-n_9\right)\Gamma
   \left(c-D/2+e-n_2-n_3-n_5-n_6-n_8-n_9\right)
\end{align}

\begin{align}
   H_{1664} &= w_1^{-b-e+D/2} w_5^{-d} w_8^{-c}
   \sum_{n_1=0}^{\infty} ... \sum_{n_9=0}^{\infty}
    \frac{(-w_2/w_8)^{n_1}}{n_1!}
    \frac{(-w_3/w_5)^{n_2}}{n_2!}
    \frac{(-w_1 w_4/w_5)^{n_3}}{n_3!}
    \nonumber \\
    & \times
    \frac{(-w_1 w_6/(w_5 w_8))^{n_4}}{n_4!}
    \frac{(-w_1 w_7/w_8)^{n_5}}{n_5!}
    \frac{(-w_9)^{n_6}}{n_6!}
    \frac{(-w_1/w_8)^{n_7}}{n_7!}
    \frac{(-w_1/w_5)^{n_8}}{n_8!}
    \frac{(-w_1)^{n_9}}{n_9!}    
        \nonumber \\
    & \times
        \Gamma \left(c+n_1+n_4+n_5+n_7\right) \Gamma \left(c+d-D/2+e+f+n_1+n_2+n_4-n_6+n_7+n_8\right)
        \nonumber \\
    & \times
        \Gamma \left(b-D/2+e-n_3-n_4-n_5-n_7-n_8-n_9\right) \Gamma
   \left(-c-d+D/2-e-n_1-n_2-n_4+n_6+n_9\right) 
        \nonumber \\
    & \times\Gamma \left(d+n_2+n_3+n_4+n_8\right) \Gamma
   \left(-b-c-d+D-e-f+n_3+n_5+n_6+n_9\right)
\end{align}

\begin{align}
   H_{1675} &= w_1^{-b-e+D/2} w_5^{-d} w_8^{-c}
   w_9^{c+d+e+f-D/2}
   \sum_{n_1=0}^{\infty} ... \sum_{n_9=0}^{\infty}
    \frac{(-w_2 w_9/w_8)^{n_1}}{n_1!}
    \frac{(-w_3 w_9/w_5)^{n_2}}{n_2!}
    \frac{(-w_1 w_4/w_5)^{n_3}}{n_3!}
    \nonumber \\
    & \times
    \frac{(-w_1 w_6 w_9/(w_5 w_8))^{n_4}}{n_4!}
    \frac{(-w_1 w_7/w_8)^{n_5}}{n_5!}
    \frac{(-w_1 w_9/w_8)^{n_6}}{n_6!}
    \frac{(-w_1 w_9/w_5)^{n_7}}{n_7!}
    \frac{(-w_9)^{n_8}}{n_8!}
    \frac{(-w_1)^{n_9}}{n_9!}    
        \nonumber \\
    & \times
        \Gamma \left(c+n_1+n_4+n_5+n_6\right) \Gamma \left(d+n_2+n_3+n_4+n_7\right)
    \nonumber \\
    & \times
        \Gamma \left(f+n_6+n_7+n_8+n_9\right) \Gamma
   \left(-b+D/2+n_1+n_2+n_3+n_4+n_5+n_6+n_7+n_8+n_9\right)
    \nonumber \\
    & \times    
        \Gamma \left(b-D/2+e-n_3-n_4-n_5-n_6-n_7-n_9\right)
        \nonumber \\
    & \times    
        \Gamma
   \left(-c-d+D/2-e-f-n_1-n_2-n_4-n_6-n_7-n_8\right)
\end{align}

\begin{align}
   H_{2062} &= w_1^{-b-e+D/2} w_3^{-c-d-e+D/2} w_5^{c-D/2+e} w_8^{-c}
   \sum_{n_1=0}^{\infty} ... \sum_{n_9=0}^{\infty}
    \frac{(-w_2 w_5/(w_3 w_8))^{n_1}}{n_1!}
    \frac{(-w_1 w_4/w_5)^{n_2}}{n_2!}
    \nonumber \\
    & \times
    \frac{(-w_1 w_6/(w_3 w_8))^{n_3}}{n_3!}
    \frac{(-w_1 w_7/w_8)^{n_4}}{n_4!}
    \frac{(-w_3 w_9/w_5)^{n_5}}{n_5!}
    \frac{(-w_3/w_5)^{n_6}}{n_6!}
    \frac{(-w_1/w_8)^{n_7}}{n_7!}
    \frac{(-w_1/w_5)^{n_8}}{n_8!}
        \nonumber \\
    & \times
    \frac{(-w_1 w_3/w_5)^{n_9}}{n_9!}
    \Gamma \left(c+n_1+n_3+n_4+n_7\right)
        \nonumber \\
    & \times
    \Gamma \left(f+n_6+n_7+n_8+n_9\right) \Gamma \left(b-D/2+e-n_2-n_3-n_4-n_7-n_8-n_9\right)
        \nonumber \\
    & \times
    \Gamma
   \left(-c+D/2-e-n_1+n_2+n_5+n_6+n_8+n_9\right) \Gamma \left(c+d-D/2+e+n_1+n_3-n_5-n_6-n_9\right)
        \nonumber \\
    & \times
   \Gamma
   \left(-b-c-d+D-e-f+n_2+n_4+n_5+n_9\right)
\end{align}

\begin{align}
   H_{2442} &= w_1^{-b+D/2-e} w_2^{-c+D/2-e} w_3^{-d} w_8^{-D/2+e}
   \sum_{n_1=0}^{\infty} ... \sum_{n_9=0}^{\infty}
    \frac{(-w_1 w_2 w_4/(w_3 w_8))^{n_1}}{n_1!}
    \nonumber \\
    & \times
    \frac{(-w_2 w_5/(w_3 w_8))^{n_2}}{n_2!}
    \frac{(-w_1 w_6/(w_3 w_8))^{n_3}}{n_3!}
    \frac{(-w_1 w_7/w_8)^{n_4}}{n_4!}
    \frac{(-w_2 w_9/w_8)^{n_5}}{n_5!}
    \frac{(-w_2/w_8)^{n_6}}{n_6!}
    \frac{(-w_1/w_8)^{n_7}}{n_7!}
        \nonumber \\
    & \times
    \frac{(-w_1 w_2/(w_3 w_8))^{n_8}}{n_8!}
    \frac{(-w_1 w_2/w_8)^{n_9}}{n_9!}
    \Gamma \left(d+n_1+n_2+n_3+n_8\right)
        \nonumber \\
    & \times
     \Gamma \left(f+n_6+n_7+n_8+n_9\right) \Gamma \left(D/2-e+n_1+n_2+n_3+n_4+n_5+n_6+n_7+n_8+n_9\right)
        \nonumber \\
    & \times
    \Gamma \left(b-D/2+e-n_1-n_3-n_4-n_7-n_8-n_9\right) \Gamma \left(c-D/2+e-n_1-n_2-n_5-n_6-n_8-n_9\right)
        \nonumber \\
    & \times
     \Gamma
   \left(-b-c-d+D-e-f+n_1+n_4+n_5+n_9\right)
\end{align}

\end{document}